\documentclass[12pt]{iopart}

\usepackage{ctable}
\usepackage{graphicx}
\usepackage{amssymb}
\usepackage{amsbsy}
\expandafter\let\csname equation*\endcsname\relax
\expandafter\let\csname endequation*\endcsname\relax
\usepackage{amsmath}
\usepackage{setspace}

\begin{document}

\title[]{Gate fidelity comparison in semiconducting spin qubit implementations affected by control noises}

\author{E. Ferraro$^1$, M. Fanciulli$^{1,2}$ and M. De Michielis$^1$}
\ead{marco.demichielis@mdm.imm.cnr.it}
\address{$^1$ CNR-IMM Unit of Agrate Brianza, Via C. Olivetti 2, 20864 Agrate Brianza (MB), Italy}
\address{$^2$ Dipartimento di Scienza dei Materiali, Universit\`a degli Studi di Milano-Bicocca, Via Cozzi 55, 20125 Milano, Italy}

\vspace{10pt}

\begin{abstract}
A comparison of gate fidelities between different spin qubit types defined in quantum dots and a donor under different control errors is reported. We studied five qubit types, namely the quantum dot spin qubit, the double quantum dot singlet-triplet qubit, the double quantum dot hybrid qubit, the donor qubit and the quantum dot spin-donor qubit. For each one, we derived analytical time sequences that realize single qubit rotations along the principal axis of the Bloch sphere. We estimated the effects of control errors on the gate fidelity by using a Gaussian noise model. Then we compared the gate fidelities among qubit implementations due to pulse timing errors by using a realistic set of values for the error parameters of control amplitudes. 
\end{abstract}

\section{Introduction}
In the framework of solid state physics a rich exploitable platform for universal quantum computation, as witnessed by several experimental \cite{Shulman-2012,Veldhorst-2014,Pla-2012,Maune-2012,Bluhm-2011,Tyryshkin-2012} and theoretical \cite{RuiLi-2012,Coish-2005,Shen-2000} proposals, is represented by the confinement of electron spins in host semiconducting materials. The confinement is achievable following different routes, from electrostatically or self-assembled quantum dots (QDs) \cite{Morton-2011,Veldhorst-2014,Kawakami-2014} to donor spins in solid matrices \cite{Klymenko-2015,Gamble-2015,Saraiva-2015} or a combination of them \cite{Urdampilleta-2015,Pica-2016,Harvey-2017}. 

QDs are also known as artificial atoms due to the fact that the confinement of electrons or holes in a semiconductor nanostructure gives birth to a new potential profile with discrete energy levels. Such discretization is created adopting external electrostatic potentials applied to metallic gates (electrostatically defined QD) or in alternative by means of an appropriate growth process of the nanostructure that creates the confinement potentials (self-assembled QD). The main differences between the two implementations lie in the different potential profile, that confines only one type of carrier when electrostatic QDs arise whereas it confines both electrons and holes in self-assembled QDs. Other differences are in the working temperature that is lower for the electrostatic QD (lower than 1K) with respect the self-assembled (around 4K) and in the experimental control of the QD that is prevalently electrical for the first and optical for the second. Next to these spin architectures, the proposal by Kane \cite{Kane-1998} suggests a qubit implementation based on the nuclear spin of donors in semiconducting host materials, such as phosphorus in silicon. The phosphorus is a shallow donor that creates in silicon an additional energy level under the conduction band, a level that can host an electron at low temperature. Also the spin of this bounded electron can be used as an additional qubit holder. 

The reasons that make semiconductor nanostructure-based qubits an attractive scenario for technological applications are due to their relatively long coherence times, the easy manipulation, fast gate operations and potential for scaling \cite{Loss-1998,DiVincenzo-2000,Taylor-2005,Laird-2010}. The semiconducting materials adopted range from III-V compounds such as GaAs to IV group materials such as Si and Ge. While the first are affected by the unavoidable hyperfine interaction due to the nuclear spin, the group IV materials present nuclear free isotopes that allow to overcome this source of decoherence. In addition the interest towards silicon qubits is immediately linked to the integrability with the already existing CMOS (complementary metal oxide semiconductor) infrastructure of the microelectronics industry. From the point of view of the qubit type several proposal are presented in the literature based on single \cite{Loss-1998}, double \cite{Taylor-2005,Levy-2002,Petta-2005} and triple \cite{DiVincenzo-2000} QDs and analogously for the donor scenario \cite{Kane-1998,Pla-2012,Pla-2013}. 

\begin{figure}[htbp]
	\begin{center}
	\includegraphics[width=1\textwidth]{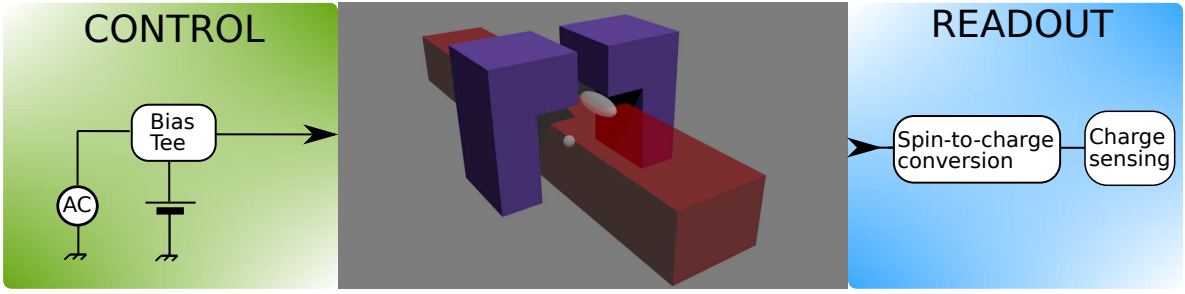}
	\caption{Sketch of the device where the five qubit types can be held. The semiconductor nanowire is highlighted in red whereas the metal gates, shown in purple, are insulated from the nanowire by a dielectric. Some exemplificative spatial distributions of electrons confined in electrostatically defined QDs (ellipsoid) and in donors (sphere) are shown in white. On the sides, control and readout blocks are sketched: in the control block, DC and AC signals are included whereas in the readout block the main steps for qubit readout are reported. The details about the external magnetic field and the microwave needed to control the qubits as well as the readout steps have been reported in the following section.}\label{Fig:Device} 
	\end{center}
\end{figure}

The five qubit types we focus on are: the quantum dot spin qubit (SQ), the double quantum dot singlet-triplet qubit (STQ), the double quantum dot hybrid qubit (HQ), the donor qubit (DQ) and the quantum dot spin-donor qubit (SDQ). Those five qubits are investigated due to the possibility, not already fully explored by experiments, to be held in a common nanoscaled device shown in Fig. \ref{Fig:Device}, where a semiconductor nanowire (in red) is electrically insulated from two metal gates (in purple) by a dielectric. Depending on the type of qubit, QDs can be electrostatically induced in semiconductor regions close to the upper corners by the metal gates and, eventually, the donor can be positioned in the semiconductor volume below the gates. 

For every qubit type the effective Hamiltonian model, the analytical sequence times that realize the single qubit operations on the Bloch sphere and the fidelity gate analysis are derived and examined putting them in comparison.

We adopt the entanglement fidelity to test the resilience of the quantum gates with respect to different disturbance sources. The ideal realization of quantum gates is deeply influenced by the unavoidable environmental noise due to different sources of disturbance that cause decoherence. In order to have a homogeneous analysis for all the five qubit types, we consider two different sources of noise linked to the $z$ and $x$ contributions appearing in the effective Hamiltonian models. Every time that the qubit type is specified such abstract controls become physical entities giving a real meaning to the type of disturbance and a measure of it.

The paper is organized as follows. Sect. 2 is devoted to the presentation of the effective Hamiltonian models for the five qubit types and to the derivation of the analytical time sequences that realize single qubit rotations on the Bloch sphere. Sect. 3 defines the global framework for all the implementations under investigation. Sect. 4 contains the main results about the single qubit gate fidelities while Sect. 5 gives a comparison among all the five qubit types. Finally in Sect. 6, concluding remarks are summarized.

\section{Spin qubit types: Hamiltonian models and single qubit operations}
\label{Sec:SpinQubitTypes}
This Section is devoted to the presentation of the effective Hamiltonian models and to the derivation of the analytical time gate sequences that realize single qubit operations for the five spin qubit types under investigation. 

The fundamental single qubit operations that are derived are the rotations on the Bloch sphere about the $z$ axis $R_{z}(\theta)$ and about the $x$ axis $R_{x}(\phi)$ whose matrix representations are

\begin{equation}\label{rotz}
R_{z}(\theta)=\left(\begin{array}{cc} 
e^{-i\theta/2} & 0  \\
0 & e^{i\theta/2} 
\end{array}\right)
\end{equation}
and
\begin{equation}\label{rotx}
R_{x}(\phi)=\left(\begin{array}{cc} 
\cos(\phi/2) & -i \sin(\phi/2)  \\
-i \sin(\phi/2) & \cos(\phi/2)
\end{array}\right).
\end{equation}

The condition to be satisfied in order to obtain the single qubit gate desired is that for each type the unitary evolution operator 
\begin{equation}\label{U}
U(t)=e^{-\frac{i}{\hbar}Ht}
\end{equation} 
coincides with the rotation matrices $R_{z}(\theta)$ and $R_{x}(\phi)$ when the Hamiltonian models are specified case by case. The results will be analytical time sequences that we are going to present for each type in the next subsections.
 
\subsection{Quantum dot spin qubit}
The SQ is realized confining the spin of a single electron in a QD. The logical qubit basis is simply defined by the two spin eigenstates $|0\rangle\equiv|\!\uparrow\rangle$ and $|1\rangle\equiv|\!\downarrow\rangle$, that correspond to the angular momentum states with $S=\frac{1}{2}$, $S_z=\frac{1}{2}$ and $S=\frac{1}{2}$, $S_z=-\frac{1}{2}$ respectively. The Hamiltonian model is given by
\begin{equation}\label{HSS}
H=\frac{\hbar}{2}\omega_z\sigma_z+\hbar\Omega_x\cos(\omega t)\sigma_x
\end{equation}
where $\sigma_{z(x)}$ is the Pauli operator, $\hbar\omega_z=g_e\mu_BB_0$ is the Zeeman energy associated to the constant magnetic field in the $z$ direction $B_0$ with $g_e$ the electron g-factor and $\mu_B$ the Bohr magneton and $\hbar\Omega_x=g_e\mu_BB_1/2$ where $B_1$ is the oscillating magnetic field and $\omega$ is its angular frequency.

Universal qubit control of the spin qubit is achieved adopting magnetic fields pulses. The two spin eigenstates $|0\rangle\equiv|\!\!\uparrow\rangle$ and $|1\rangle\equiv|\!\!\downarrow\rangle$ are separated by an energy $\Delta E$. They can be coherently manipulated using resonant microwave pulses of frequency $\omega$. From a practical point of view the approaches implemented are based on: electron spin resonance (ESR) techniques using local AC magnetic fields \cite{Morton-2011}, application of global magnetic fields for local manipulation or through all-electrical manipulation via AC electric fields in a magnetic field gradient \cite{Veldhorst-2014}. The first step in the control and manipulation of the QD spin qubit is represented by the initialization, which requires a measurement on the single spin. Spin states in QDs are deeply studied by measuring the average signal from a large ensemble of electron spins \cite{Ciorga-2001}. In contrast single-shot measurement, where it is tested if the spin orientation is parallel or anti-parallel to the external magnetic field, is more appealing \cite{Elzerman-2004}. Finally the read-out, in which the state of the qubit is determined after implementing the algorithm, can rely on spin-to-charge conversion, where the tunneling of a single electron can be measured through an integrated charge sensor or can exploit RF reflectometric techniques, using gate electrodes coupled to off-chip lumped-element resonators \cite{Gonzalez-2015}.
Although the manipulation requires sophisticated techniques involving magnetic fields or gradient of them, such spin qubit has a great advantage represented by long coherence times of the order of milliseconds \cite{Kawakami-2014}.

The Hamiltonian in Eq. (\ref{HSS}) is explicitly time-dependent and prevents itself from an analytical solution. However, it is possible to transform the Hamiltonian into a rotating frame, which rotates at the angular frequency $\omega$. Under the rotating wave approximation (RWA), where the fast oscillating terms (2$\omega$) of the Hamiltonian in the rotating frame are neglected, the total Hamiltonian finally reads 
\begin{equation}\label{HSS2}
H=H_z+H_x= \frac{\hbar}{2}(\omega_z-\omega)\sigma_z + \frac{\hbar}{2}\Omega_x\sigma_x, 
\end{equation}
and the Larmor angular frequency $\omega_z$ and the amplitude $\Omega_x$ of the AC local magnetic field are adopted as input controls. The rotation matrix $R_z(\theta)$ is obtained by a one step sequence with time
\begin{equation}
t_z(\theta) = \frac{\theta}{\Delta\omega_z}.
\end{equation}
where $\Delta\omega_z\equiv\omega_z-\omega$.

The rotation matrix $R_x(\phi)$ is obtained by injecting the Hamiltonian (\ref{HSS2}) into Eq. (\ref{U}) and equaling the result with Eq. (\ref{rotx}). The resulting analytical expressions for the time realizing the single qubit rotation along $x$ axis via ESR (with $\omega=\omega_z$) is given by
\begin{equation}
t_x(\phi) = \frac{\phi}{\Omega_x}.
\end{equation}

\subsection{Double quantum dot singlet-triplet qubit}
The STQ is created from two electrons, ideally spatially separated, in two QDs. The logical states are defined by a superposition of two-particle spin singlet and triplet states, that are $|0\rangle\equiv|S\rangle$ and $|1\rangle\equiv|T_0\rangle$, where each QD is occupied with one electron. $|S\rangle$, $|T_0\rangle$ and $|T_{\pm}\rangle$ are respectively the singlet and triplet states of a pair of electrons given by
\begin{eqnarray}\label{st1}
&|S\rangle=\frac{1}{\sqrt{2}}(|\!\uparrow\!\downarrow\rangle-|\!\downarrow\!\uparrow\rangle), \quad |T_0\rangle=\frac{1}{\sqrt{2}}(|\!\uparrow\!\downarrow\rangle+|\!\downarrow\!\uparrow\rangle),\nonumber\\
&|T_-\rangle=|\!\downarrow\!\downarrow\rangle, \quad |T_+\rangle=|\!\uparrow\!\uparrow\rangle.
\end{eqnarray}
It is an external magnetic field that removes the $|T_-\rangle$ and $|T_+\rangle$ branches. 
Qubit state rotations are performed acting on the QDs energy detuning and causing a change in the exchange coupling $J$ between the two electrons. In addition a local magnetic field gradient is necessary to achieve arbitrary qubit rotations. Note that the energy detuning can also be created with hyperfine coupling or a difference in spin orbit coupling but in this paper we will focus on detuning induced by electrical potentials. The Hamiltonian model 
\begin{equation}\label{singlet-triplet}
H=\frac{1}{2}\Delta E_z(\sigma_{1}^z-\sigma_{2}^z)+\frac{1}{4}J\boldsymbol{\sigma}_{1}\cdot\boldsymbol{\sigma}_{2}
\end{equation}
contains the exchange interaction between the two electrons, described by the Pauli matrices $\boldsymbol{\sigma}_1$ and $\boldsymbol{\sigma}_2$, through the coupling constant $J$, beyond the Zeeman term. It corresponds to a magnetic field gradient between the QDs adopted for the single qubit control, that is $\Delta E_z=\frac{1}{2}(E_1^z-E_2^z)$. The exchange coupling $J$ is derived from a generalized Hubbard model adopting a standard procedure based on the Schrieffer-Wolff transformation \cite{Li-2012} and it is equal to
\begin{equation}
J=\frac{4(t-J_t)^2}{U-U'-|\Delta\varepsilon|}-2J_e.
\end{equation}
The parameters involved are: the single-electron tunneling $t$ across the double QD, the single-electron tunneling in the presence of a second electron $J_t$, the intradot (interdot) Coulomb repulsion U (U'), the interdot bias $\Delta\varepsilon$, that is the single electron ground-orbital energy difference between the two dots and the direct exchange interaction $J_e$ of the two electrons across the double QD.

Initialization of STQ is achieved by loading an electron in the single occupied QD, going from $(1,0)$ to $(2,0)$ charge configuration, to obtain a singlet spin state. Readout of STQ have been successfully achieved using the Pauli spin blockade \cite{Zwanenburg-2013}. This qubit type allows fast readout and fast manipulation, however experimentally the big challenge is represented by the creation of the local magnetic gradient \cite{XianWu-2014,Barnes-2016,Pioro-2008,Maune-2012,Gonzalez-2015}. A valid strategy to overcome such task is represented by the use of a micromagnet in close proximity \cite{Barnes-2016,Pioro-2008,XianWu-2014}.

In the qubit basis $\{|S\rangle, |T_0\rangle\}$ the Hamiltonian can be recast in a compact form 
\begin{equation}
H= \begin{pmatrix}
 -\frac{3}{4}J & \Delta E_z &\\
 \Delta E_z & \frac{1}{4}J &\end{pmatrix}.
\end{equation}
In order to obtain the rotation matrix $R_z(\theta)$, let's consider the two steps case where the resulting matrix is the product of the evolution operators $U_{J}(t_{J})\cdot U_z(t_z)$ with $H_{J}=\frac{1}{4}J\boldsymbol{\sigma}_{1}\cdot\boldsymbol{\sigma}_{2}$ and $H_z=\frac{1}{2}\Delta E_z(\sigma_{1}^z-\sigma_{2}^z)$. In the expressions for $H_J$ we made the assumption that $J\gg\Delta E_z$, with $J(\epsilon)$ varying between $J$ and $0$, consequently the term proportional to $\Delta E_z$ in the Hamiltonian is negligible. After carrying out the calculations it is found that
\begin{equation} 
U_{J}(t_{J})\cdot  U_z(t_z) = 
\left(\begin{array}{cc}
\alpha \cos(\gamma) & -i \alpha \sin(\gamma)\\
-i \beta \sin(\gamma)  & \beta \cos(\gamma)  
\end{array}\right) 
\end{equation}
where
\begin{eqnarray} \label{Eq:alphaBetaGamma}
\alpha &=& \exp(i2\pi 3/4 Jt_J/h)\nonumber \\
\beta &=& \exp(-i2\pi 1/4 Jt_J/h)\nonumber \\
\gamma &=& 2\pi \Delta E_z t_z/h.
\end{eqnarray}
By imposing 
\begin{equation} \label{Eq:matrixProduct2}
\left(\begin{array}{cc}
\alpha \cos(\gamma) & -i \alpha \sin(\gamma)  \\
-i \beta \sin(\gamma)  & \beta \cos(\gamma)  
\end{array}\right) 
= \left(\begin{array}{cc} 
e^{-i\theta/2} & 0  \\
0 & e^{i\theta/2} 
\end{array}\right)
\end{equation}
and that the time must be non-negative and as short as possible, we obtain
\begin{eqnarray}
&t_z=\frac{n}{2}\frac{h}{\Delta E_z}\nonumber\\
&t_{J}(\theta)=\left(-\frac{\theta}{2\pi}+n\right)\frac{h}{J}
\end{eqnarray}
with $n=1$ and $0\leq\theta<2\pi$.
 
Analogously for the $R_x(\phi)$ gate we have to impose 
\begin{equation} \label{Eq:matrixProduct3}
\left(\begin{array}{cc}
\alpha \cos(\gamma) & -i \alpha \sin(\gamma)  \\
-i \beta \sin(\gamma)  & \beta \cos(\gamma)  
\end{array}\right) 
= \left(\begin{array}{cc} 
\cos(\phi/2) & -i\sin(\phi/2)  \\
-i\sin(\phi/2) & \cos(\phi/2) 
\end{array}\right)
\end{equation}
that is $\alpha=\beta=1$ and $\gamma=\phi/2$. It follows that 
\begin{eqnarray}
&t_z(\phi)=\left(\frac{\phi}{4\pi}+n\right)\frac{h}{\Delta E_z}\nonumber\\
&t_{J}=0
\end{eqnarray}
that is minimum when $n=0$.

\subsection{Double quantum dot hybrid qubit}
The HQ owes its name to the fact that is an \emph{hybrid} of spin and charge \cite{Shi-2012}. It is composed by two QDs in which three electrons have been confined with all-electrical control via gate electrodes. The logical states coded using the $S=\frac{1}{2}$ and $S_z=\frac{1}{2}$ three electrons subspace, have been defined by adopting combined singlet and triplet states of a pair of electrons occupying one dot with the states of the single electron occupying the other. The logical states have been expressed by $|0\rangle\equiv|S\rangle|\!\uparrow\rangle$ and $|1\rangle\equiv\sqrt{\frac{1}{3}}|T_0\rangle|\!\uparrow\rangle-\sqrt{\frac{2}{3}}|T_+\rangle|\!\downarrow\rangle$ where $|S\rangle$, $|T_0\rangle$ and $|T_{\pm}\rangle$ are defined in Eq. (\ref{st1}). The effective Hamiltonian model involving only exchange interaction terms among couples of electrons for a single and two qubits was derived in Ref. \cite{Ferraro-2014} and in Ref. \cite{Ferraro-2015-qip}, respectively. HQ has been deeply investigated and constantly developed as witnessed by several recent papers \cite{DeMichielis-2015,Ferraro-2015-prb,Ferraro-2017,Ferraro-QIP2018,Rotta-2016}. For the single HQ the effective Hamiltonian is equal to 
\begin{equation}\label{Hy}
H=\frac{1}{2}E_z(\sigma_{1}^z+\sigma_{2}^z+\sigma_{3}^z)+\frac{1}{4}J'\boldsymbol{\sigma}_{1}\cdot\boldsymbol{\sigma}_{2}+\frac{1}{4}J_{1}\boldsymbol{\sigma}_{1}\cdot\boldsymbol{\sigma}_{3}+\frac{1}{4}J_{2}\boldsymbol{\sigma}_{2}\cdot\boldsymbol{\sigma}_{3},
\end{equation} 
where the effective coupling constants are given in Ref. \cite{Ferraro-2014}.

In the initialization process, all the variables are regulated through appropriate external electric and magnetic fields driving the qubit in the state corresponding to the $|0\rangle$ logical state. After the initialization, it is possible to lead the desired logical gates through operations that are generally described by unitary matrices. In order to inject electrons in the QDs a reservoir as source of electrons near the double QD is required. The height of the energy barrier between the reservoir itself and the double QD is controlled through an electrostatic gate. A charge sensor enables the readout of the spin state of electrons confined in the doubly occupied QD. A single-electron transistor (SET) can be used to electrostatically sense the spin state of the electrons. 
More in detail, when readout of the qubit starts, tunneling is allowed from the doubly occupied QD to some reservoir by a reduction in the interposed electrostatic barrier. When the electron pair is in a singlet state the corresponding wavefunction is more confined and the tunneling rate to the reservoir is lower than that of the triplet state, which has a broader wavefunction. When the electron tunnels, the electrostatic potential landscape changes and so does the current passing through the electrostatically coupled SET. The measurement of the time interval between the read out signal and the current variation in the SET is supposed to reveal the spin state of the electron pair.

The key advantage of this qubit type is that the manipulation of the qubit is all electrical and very fast. The coherence times of tens of nanoseconds \cite{Koh-2012,Kim-2012,Kim-2015}, shorter than that of the qubit types previously introduced, have been recently enhanced 10 fold to the 177 ns level \cite{Thorgrimsson-2017}. 

Explicit calculations of the matrix elements of the Hamiltonian (\ref{Hy}) in the logical basis give
\begin{equation}\label{effmatrix}
H= \begin{pmatrix}
 -\frac{E_z}{2}-\frac{3}{4}J' &\;\; -\frac{\sqrt{3}}{4}(J_1-J_2) \\
 -\frac{\sqrt{3}}{4}(J_1-J_2) &\;\; -\frac{E_z}{2}+\frac{1}{4}J'-\frac{1}{2}(J_1+J_2) \end{pmatrix}.
\end{equation}
The sequence that realizes $R_z(\theta)$ is composed by the product of three steps $U_{J_1}(t_{J_1})\cdot U_{J'}(t_{J'})\cdot U_{J_2}(t_{J_2})$ where the evolution operators are calculated starting from the Hamiltonians $H_{J_i}=\frac{1}{2}E_z(\sigma_{1}^z+\sigma_{2}^z+\sigma_{3}^z)+\frac{1}{4}J'\boldsymbol{\sigma}_{1}\cdot\boldsymbol{\sigma}_{2}+\frac{1}{4}J_{i}\boldsymbol{\sigma}_{i}\cdot\boldsymbol{\sigma}_{3}$ with $i=1,2$ and $H_{J'}=\frac{1}{2}E_z(\sigma_{1}^z+\sigma_{2}^z+\sigma_{3}^z)+\frac{1}{4}J'\boldsymbol{\sigma}_{1}\cdot\boldsymbol{\sigma}_{2}$. We put the control parameters to $\max(J_1) = \max(J_2) = J^{max}$. We made the assumption of ideal no crosstalk between $J_1$ and $J_2$ by considering a large energy gap between the ground and excited energy levels in the doubly occupied dot and an even higher energy gap between the ground and excited level in the singly occupied dot. In this case, $J_1$ can be switched on and off without affecting $J_2$ by aligning the energy levels of the ground state of the doubly occupied quantum dot with the ground level in the singly occupied quantum dot and lowering the energy barrier between them. Therefore, $J_1$ is maximized whereas $J_2$ is not perturbed because the excited state of the singly occupied quantum dot is at a higher energy. On the contrary $J'$ is set as a constant equal to $J^{max}/2$ so it is never at zero.

The results of the calculations lead after algebraic manipulations to the following analytical expressions for the times:
\begin{eqnarray}\label{rz}
&t_{J_1}(\theta)=\frac{1}{C}\left[\frac{\theta}{\pi}A+sign\left(\frac{2\pi}{3}-\theta\right)B\right]\frac{h}{J^{max}}\nonumber\\
&t_{J_2}(\theta)=t_{1}(\theta)\nonumber\\
&t_{J'}(\theta)=\left(2-\frac{\theta}{\pi}\right)\frac{h}{J^{max}}
\end{eqnarray}
where
\begin{eqnarray}\label{coeff}
&A=\frac{E_z}{2}+\frac{1}{8}J^{max}\nonumber\\
&B=-E_z+\frac{1}{4}J^{max}\nonumber\\
&C=E_z+\frac{3}{4}J^{max}.
\end{eqnarray}
The $R_x(\phi)$ sequence is instead composed by the product of two steps $U_{J_1}(t_{J_1})\cdot U_{J_2}(t_{J_2})$ where the times are given by:
\begin{eqnarray}\label{rx}
&t_{J_1}(\phi)=\left(\frac{n}{C}-\frac{1}{\sqrt{3}}\frac{\phi}{2\pi}\frac{1}{J^{max}}\right)h\nonumber\\
&t_{J_2}(\phi)=\left(\frac{n}{C}+\frac{1}{\sqrt{3}}\frac{\phi}{2\pi}\frac{1}{J^{max}}\right)h,
\end{eqnarray}
with
\begin{equation}
n=\left\lceil \frac{C}{J^{max}}\frac{1}{\sqrt{3}}\frac{\phi}{2\pi} \right\rceil.
\end{equation}

\subsection{Donor qubit}
In close analogy to the SQ, an alternative kind of qubit to store and manipulate the quantum information is represented by the DQ. It is implemented on the spin of an electron bound to a donor and the qubit basis is given by the single electron spin $|0\rangle\equiv|\!\uparrow\rangle$ and $|1\rangle\equiv|\!\downarrow\rangle$ logical states. The spin Hamiltonian describing the DQ common for group V donors in silicon is given by \cite{Kane-1998,Pla-2012,Mohammady-2012,Sousa-2003}
\begin{equation}\label{donor}
H=\gamma_eB_0S_z-\gamma_nB_0I_z+\Omega_x\cos(\omega t)S_x+A\mathbf{S}\cdot\mathbf{I},
\end{equation}
the first two terms is the sum of the electronic $\mathbf{S}$ and nuclear $\mathbf{I}$ spin Zeeman interactions with an external field $B_0$. The electron and nuclear gyromagnetic ratio $\gamma_e$ and $\gamma_n$ are equal respectively to $\mu_Bg_e$ and $\mu_ng_n$, where $\mu_B$ $(\mu_n)$ is the Bohr (nuclear) magneton and $g_e$ $(g_n)$ the electron (nuclear) g-factor. The term proportional to $\Omega_x$ represents the microwave control. The last term corresponds to the hyperfine coupling. The contact hyperfine interaction energy is given by $A=\frac{8}{3}\pi\mu_B g_n\mu_n|\psi(0)|^2$ where $|\psi(0)|^2$ is the probability density of the electron wavefunction evaluated at the nucleus \cite{Kane-1998}. The electron and nuclear gyromagnetic ratios as well as the hyperfine constant $A$ are measurable by estimating the magnetic field dependencies of the spin transition frequencies.

An emblematic example is constituted by a device in which implanted phosphorous donors are coupled to metal-oxide-semiconductor single-electron transistor. Electron spin resonance is used to drive Rabi oscillations in combination with Hahn echo pulse sequence. Because of the weak spin-orbit coupling, coherence times observed are longer in comparison with the previous architectures and can reach values ranging from 3 to 6 seconds \cite{Wolfowicz-2013,Morello-2010}. On the other hand the control and manipulation of the single spin \cite{Pla-2012,Pla-2013}, as in the QD spin qubit, take place using local AC or global magnetic fields. This aspect deserves special attention from a practical point of view, in order to realize large scale arrays.

Let's consider the case in correspondence to $I=1/2$ which applies for example in the case of $^{31}P$. In the high field limit, i.e. $\gamma_e B_0\gg A$, the diagonal terms of the hyperfine interaction become negligible and the ESR allowed transition are confined in two distinct subspaces, one in which the nuclear spin has down projection $|\!\Downarrow\rangle$ and the complementary in which the nuclear spin has up projection $|\!\Uparrow\rangle$. The Hamiltonian models that effectively describe the DQ in the two subspaces are expressed by
\begin{equation}\label{Hdown}
H_{\{\uparrow,\downarrow\}\otimes\Downarrow}=\frac{\hbar}{2} (\omega_{12}-\omega)\sigma_z+\frac{\hbar}{2}\Omega_x\sigma_x
\end{equation}
and
\begin{equation}\label{Hup}
H_{\{\uparrow,\downarrow\}\otimes\Uparrow}=\frac{\hbar}{2} (\omega_{34}-\omega)\sigma_z+\frac{\hbar}{2}\Omega_x\sigma_x,
\end{equation}
where $\omega_{12}=\Delta_-+\sqrt{\Delta_+^2+4a^2}-2a$, $\omega_{34}=\Delta_-+\sqrt{\Delta_+^2+4a^2}+2a$ with $\Delta{\pm}=\frac{1}{2}(\gamma_e\pm\gamma_n)B_0$ and $a=\frac{A}{4}$.

The structure of the effective Hamiltonian models (\ref{Hdown}) and (\ref{Hup}) is completely analogous to the one of the SQ (Eq. (\ref{HSS2})) and allow us to derive single qubit gates following an analytical procedure. The condition to be satisfied is that the unitary evolution operators coincide with the rotation matrices (\ref{rotz}) and (\ref{rotx}) and the resulting analytical expressions, in the subspace in correspondence to nuclear spin with down projection and adopting as input control $\Delta\omega_{12}$ and $\Omega_x$, are given for $R_z(\theta)$ by a one step sequence with time
\begin{equation}
t_z(\theta)=\frac{\theta}{\Delta\omega_{12}}.
\end{equation}
where $\Delta\omega_{12}\equiv\omega_{12}-\omega$.

$R_x(\phi)$ is obtained by a one step sequence with 
\begin{equation}
t_x(\phi)=\frac{\phi}{\Omega_x}.
\end{equation}

Analogously single qubit gates are obtainable in the subspace corresponding to the nuclear spin with up projection described by the effective Hamiltonian model (\ref{Hup}).

\subsection{Quantum dot spin-donor qubit}
The SDQ type is the analogous of the STQ qubit in which the exchange interaction between a SQ and a DQ is exploited \cite{Urdampilleta-2015, Pica-2016}, besides the hyperfine interaction between the electronic and nuclear spin of the donor. Thanks to this analogy, the effective Hamiltonian model is easily written in terms of all the angular momentum operators involved
\begin{equation}\label{SDQ}
H=\gamma_eB_0(S_{donor}^z+S_{dot}^z)-\gamma_nB_0I_z+A\mathbf{S}_{donor}\cdot\mathbf{I}+J\mathbf{S}_{donor}\cdot\mathbf{S}_{dot},
\end{equation}
where $\mathbf{S}_{donor}$ $(\mathbf{S}_{dot})$ denotes the electron spin operator of the donor (QD) and $\mathbf{I}$ is the donor nuclear spin, $\gamma_e$ ($\gamma_n$) is the electron (nuclear) gyromagnetic ratio, $B_0$ is the applied DC magnetic field and $A$ is the hyperfine coupling between the electron spin and the nuclear spin of the donor. The constant $J$ is the exchange coupling between the electron spins of the donor and of the dot.
Besides serious aspects that have to be taken into account relating to the fabrication, such qubit type assures fast readout and fast manipulation via GHz one-axis electrical control \cite{Muhonen-2014}. The most obvious obstacle in the construction of the quantum computer that exploits such type is represented by the incorporation of the donor array into the Si layer beneath the barrier layer.

This qubit type offers more than one choice in order to define the qubit and consequently the logical states necessary to perform quantum operation, the choice mainly depends on the transition energies. Single qubit operations can be performed on the donor electron spin and on the dot electron spin with a pulsed microwave field, which can be delivered locally or globally by placing devices into microwave cavities. We choose the logical basis as in Ref. \cite{Urdampilleta-2015} defined by the singlet-triplet states between the electronic spin in the QD and the electronic spin in the donor.

The structure of the effective Hamiltonian model (\ref{SDQ}) in the logical basis is completely analogous to the one of the STQ, when for example the nuclear spin has down projection 
\begin{equation}
H= \begin{pmatrix}
\frac{1}{4}\gamma_nB_0-\frac{3}{16}J & \frac{A}{16} \\
\frac{A}{16} & \frac{1}{4}\gamma_nB_0+\frac{1}{16}J \end{pmatrix}.
\end{equation}

Recognizing the input controls in the exchange coupling $J$ and in the hyperfine coupling $A$, the rotation $R_z(\theta)$ is obtainable by a two steps sequence $U_{J}(t_{J})\cdot U_A(t_A)$ where the analytical time gate sequences are given by 
\begin{eqnarray} 
&t_{J}(\theta)=\left(-\frac{\theta}{2\pi}+n\right)\frac{h}{J/4}\nonumber\\
&t_{A}=\frac{n}{2}\frac{h}{A/16}   
\end{eqnarray}
that are minimized when $n=1$. Analogously for $R_x(\phi)$ the two steps sequence is given by
\begin{eqnarray}
&t_{J}=0\nonumber\\
&t_{A}(\phi)=\left(\frac{\phi}{4\pi}+n\right)\frac{h}{A/16}.  
\end{eqnarray}
that is minimum when $n=0$.

\section{Quantum dot and donor spin qubit types in a global framework}
The five qubit types presented in the previous section have in common a compact effective Hamiltonian when expressed each in its proper logical basis $\{|0\rangle,|1\rangle\}$ given in Tab. \ref{LB}. 
\begin{table}
	\caption{\label{LB} States of the logical basis.}
	\begin{indented}
		\item[]\begin{tabular}{@{}lll}
			\br
			Qubit  & $|0\rangle$ & $|1\rangle$\\
			\mr
		    SQ & $|\!\uparrow\rangle$ & $|\!\downarrow\rangle$ \\ 
			STQ & $|S\rangle$ & $|T_0\rangle$  \\ 
			HQ & $|S\rangle|\!\uparrow\rangle$ & $\sqrt{\frac{1}{3}}|T_0\rangle|\!\uparrow\rangle-\sqrt{\frac{2}{3}}|T_+\rangle|\!\downarrow\rangle$ \\
			DQ & $|\!\uparrow\Downarrow\rangle$ & $|\!\downarrow\Downarrow\rangle$ \\
			SDQ & $|S\!\Downarrow\rangle$ & $|T_0\!\Downarrow\rangle$\\ 
			\br
		\end{tabular}
	\end{indented}
\end{table}

The effective Hamiltonian models in terms of $2\times 2$ Pauli matrices $\sigma_z$ and $\sigma_x$ and the identity operator $I_2$ are expressed by
\begin{equation}
H=\alpha_z\sigma_z+\alpha_x\sigma_x+\alpha_0 I_2, 
\end{equation}
where $\alpha_z$, $\alpha_x$ and $\alpha_0$ are given in Tab. \ref{Heff}.
\begin{table}
	\caption{\label{Heff} Coefficients of the effective Hamiltonian models.}
	\begin{indented}
		\item[]\begin{tabular}{@{}llll}
			\br
			Qubit     & $\alpha_z$ & $\alpha_x$ & $\alpha_0$\\
			\mr
		    SQ (rot. frame) & $\frac{\hbar}{2}(\omega_z-\omega)$ & $\frac{\hbar}{2}\Omega_x$ & 0\\ 
			STQ             & $-\frac{1}{2}J$ & $\Delta E_z$ & $-\frac{1}{4}J$ \\
			HQ              & $-\frac{1}{2}J'+\frac{1}{4}(J_1+J_2)$ & $-\frac{\sqrt{3}}{4}(J_1-J_2)$& $-\frac{E_z}{2}-\frac{1}{4}(J'+J_1+J_2)$\\
			DQ (rot. frame) & $\frac{\hbar}{2}(\omega_{12}-\omega)$ & $\frac{\hbar}{2}\Omega_x$ & 0\\
			SDQ             & $-\frac{1}{8}J$ & $\frac{A}{16}$  & $\frac{1}{4}\gamma_nB_0-\frac{1}{16}J$\\
			\br
		\end{tabular}
	\end{indented}
\end{table}

The control parameters to implement $R_z(\theta)$ and $R_x(\phi)$ gates for each qubit type are reported in Tab. \ref{Controls}. 
\begin{table}
	\caption{\label{Controls} External controls that realize the single qubit operations on the Bloch sphere.}
	\begin{indented}
		\item[]\begin{tabular}{@{}llll}
			\br
			Qubit & Controls for $R_z(\theta)$ & Controls for $R_x(\phi)$\\
			\mr
		        SQ & $\Delta\omega_z$ & $\Omega_x$ \\ 
			STQ & $J$, $\Delta E_z$ & $\Delta E_z$  \\ 
			HQ & $J^{max}$ & $J^{max}$\\ 
			DQ & $\Delta\omega_{12}$ & $\Omega_x$  \\ 
			SDQ & $J$, $A$ & $A$\\ 
			\br
		\end{tabular}
	\end{indented}
\end{table}

Rotations of $\pi/2$ along $x$ and $z$ axis of the Bloch sphere are chosen as reference gates for each qubit type. Table \ref{Tab:Allsequences} reports the sequences with control pulse amplitudes and times to implement the above-mentioned gates for the five qubit types. Moreover the value of the external magnetic field applied for each qubit is specified. The amplitudes of control signals, as well as the $B_0$ values, are taken from the literature as example to set realistic parameter values of each qubit model.  In this study we are making the assumption that in each gate sequence only one signal is switched on and off with ideal edges between 0 and a maximum value, fulfilling the requirement to have pulses lasting longer than a minimum time interval of 100 ps. 

\begin{table}
	\caption{\label{Tab:Allsequences} For each qubit type, external magnetic fields and control sequences with pulse type, step amplitude, step time and total time of the sequence to generate the selected gates are reported. The values of $B_0$ and the amplitudes of the control parameters are extracted from the literature.}
	\begin{indented}
		\item[]\begin{tabular}{@{}llllllll}
			\br
			Qubit & $B_0$ [T] & Operation & Step \#     & Pulse    & Step ampl. & Step time [ns] & Total time [ns]\\
			\mr	                   
			SQ & 1.2  & $R_x(\pi/2)$   & 1 & $\Omega_{x}$     & 5 MHz \cite{Zajac-2018}  & 50 & 50   \\  \cline{3-8} 
			& &  $R_z(\pi/2)$    & 1 & $\Delta \omega_z $ & 20 MHz \cite{Li-2018} & 12.5 & 12.5 \\
			\mr
			STQ & 0.03 &$R_x(\pi/2)$ & 1 & $\Delta E_{z}$ & 32 neV \cite{XianWu-2014}  & 16.15 & 16.15\\ \cline{3-8}
			& &$R_z(\pi/2)$     & 1 & $\Delta E_{z}$ & 32 neV                     & 64.62 & \\
			& &		   & 2 & $J$            & 700 neV \cite{XianWu-2014}& 4.43  & 69.05 \\
			\mr
			HQ & 0.03  & $R_x(\pi/2)$   & 1 & $J_{1}$  & 1 $\mu$eV \cite{DeMichielis-2015} & 0.38 & \\
			& &		   & 2 & $J_{2}$  & 1 $\mu$eV		              & 1.37 & 1.58 \\ \cline{3-8}		 	
			& &$R_z(\pi/2)$      & 1 & $J_{1}$  & 1 $\mu$eV		              & 3.58 & \\
			& &   		   & 2 & $J_{2}$  & 1 $\mu$eV		              & 3.58 &  \\ 
			& &		   & 3 & wait  	  & 0		                      & 6.20 &  10.36\\ 	
			\mr
			DQ & 1.5  & $R_x(\pi/2)$   & 1 & $\Omega_{x}$        & 500 kHz \cite{Muhonen-2014}  & 500 & 500 \\ \cline{3-8} 
			& & $R_z(\pi/2)$   & 1 & $\Delta\omega_{12}$ & 2 MHz \cite{Laucht-2015}& 125 & 125\\
			\mr
			SDQ & 0.3 &$R_x(\pi/2)$   & 1 & $A$ & 400 neV  \cite{Harvey-2017} & 20.68 & 20.68 \\ \cline{3-8} 
			& &$R_z(\pi/2)$     & 1 & $A$ & 400 neV                    & 82.71 & \\
			& &                 & 2 & $J$ & 100 neV  \cite{Harvey-2017}& 124.07 & 206.78\\
			\br
		\end{tabular}
	\end{indented}
\end{table}

\section{Single qubit gate infidelity}
Non idealities must be included in the model to perform a good performance analysis in real systems. We account for error sources such as time interval error (TIE) and non ideal control of the amplitude in pulse sequences for each qubit type. 
Employing the quasi-static model, the control errors are modeled as random variables with Gaussian distributions featuring zero mean and standard deviation $\sigma$ that add up to the ideal values of the corresponding control variables presented in the whole Sec. \ref{Sec:SpinQubitTypes}. The figure of merit used to estimate the disturbance effects is the entanglement fidelity $F$ \cite{Nielsen-2000,Marinescu-2012}. 
A disturbed operation $U_{d}$ affects  
\begin{equation}
F= tr [\rho^{RS} \mathbf{1}_{R} \otimes (U_{i}^{-1}U_{d})_{S} \rho^{RS} \mathbf{1}_{R} \otimes (U_{d}^{-1}U_{i})_{S}]
\end{equation}
where $U_{i}$ is the ideal time evolution and $\rho^{RS}=|\psi\rangle\langle\psi|$ with $|\psi\rangle=\frac{1}{\sqrt{2}}(|00\rangle+|11\rangle)$ represents a maximally entangled state in a double state space generated by two identical Hilbert spaces $R$ and $S$.

\subsection{Quantum dot spin qubit}
\begin{figure}[htbp]
	\begin{center}
		a)
		\includegraphics[width=0.3\textwidth]{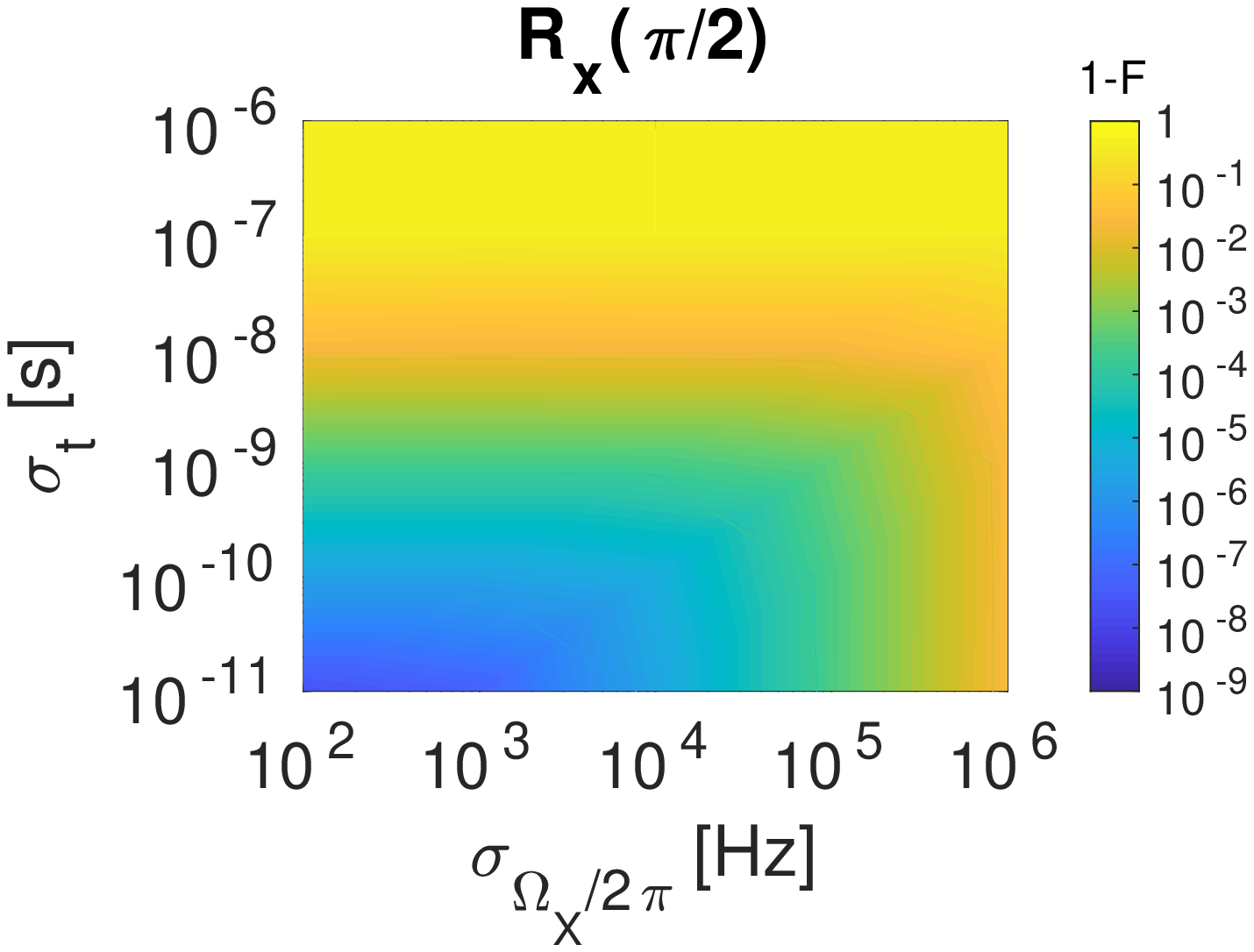}
		\includegraphics[width=0.3\textwidth]{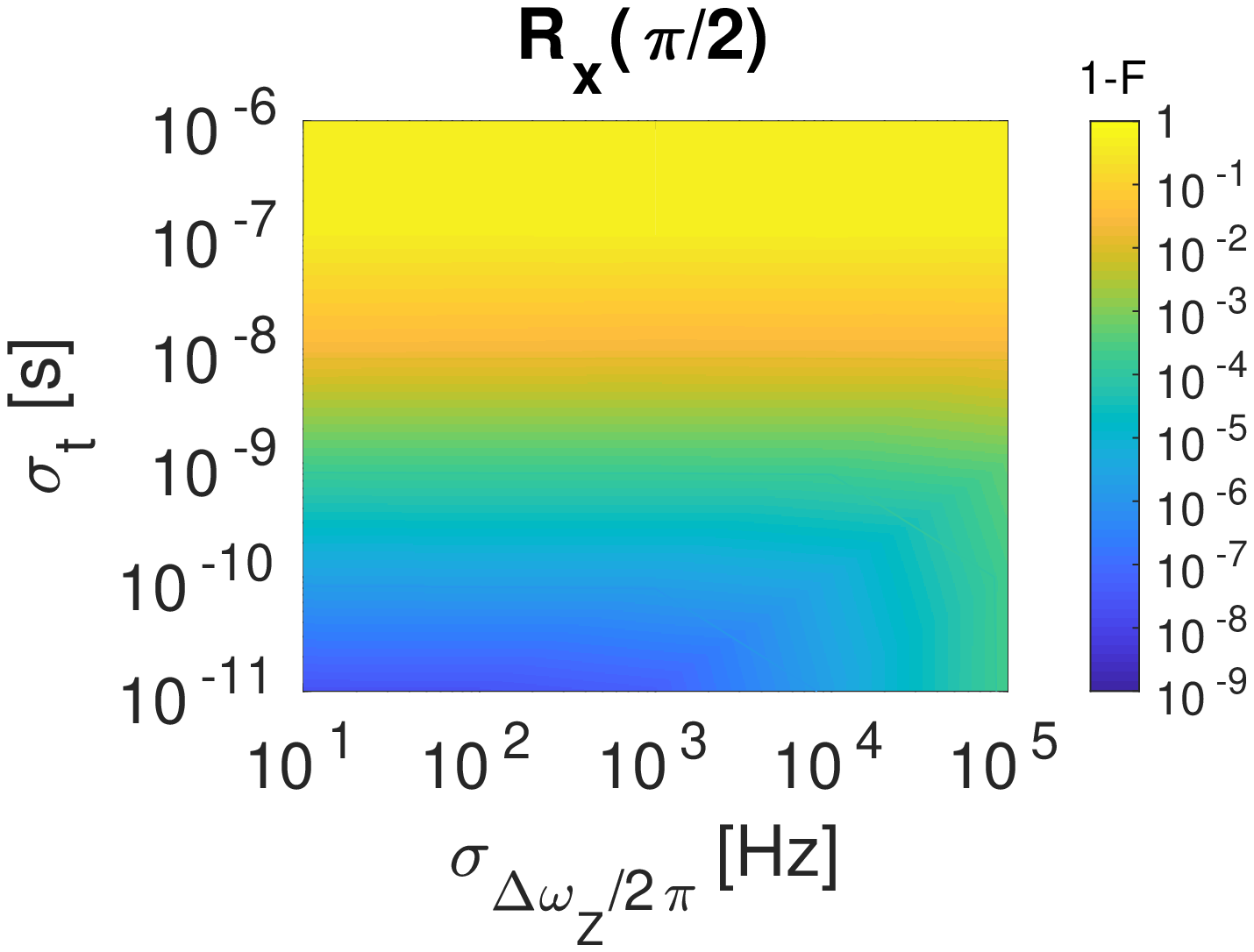}
		\includegraphics[width=0.3\textwidth]{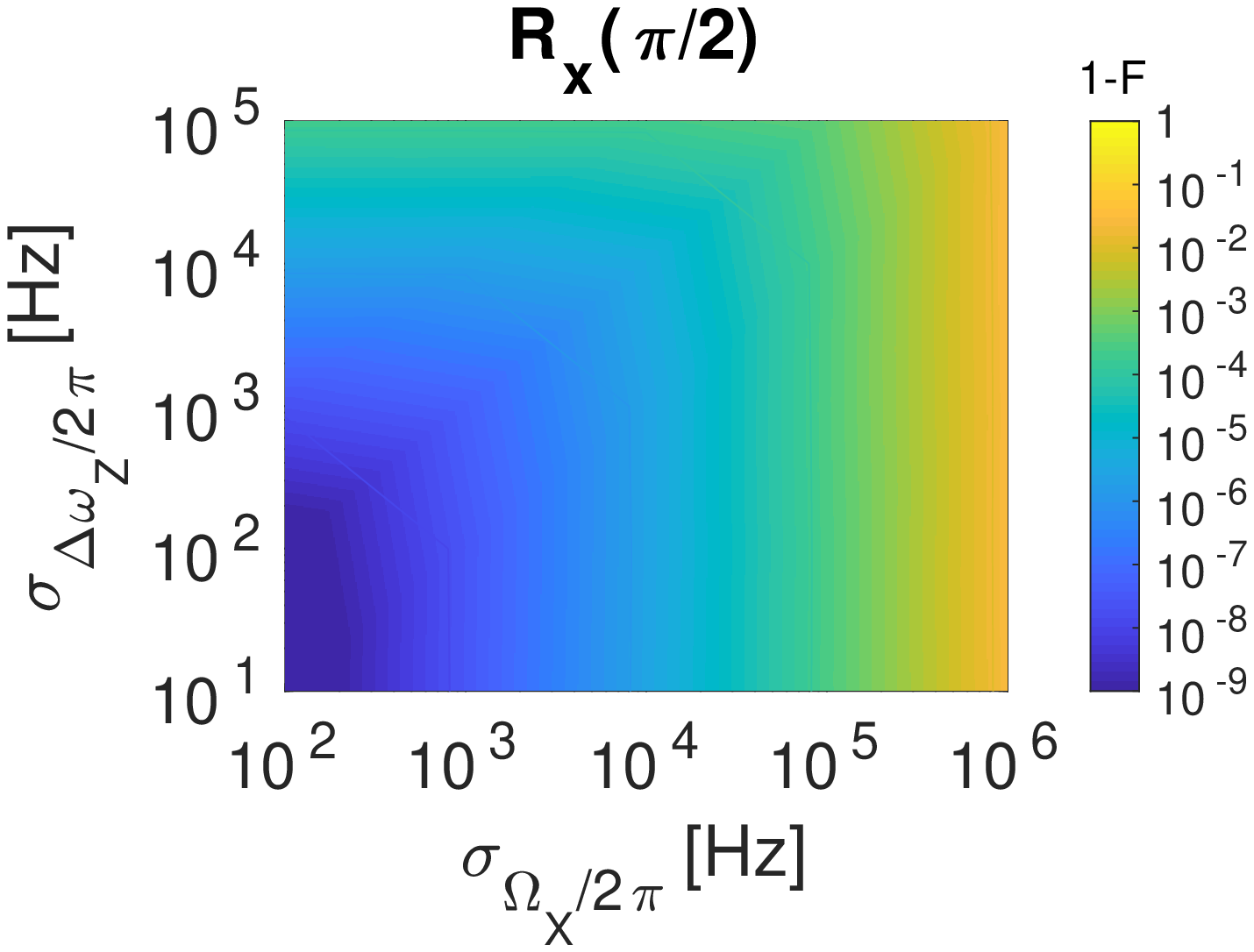}
	\end{center}
	\begin{center}
		b)
		\includegraphics[width=0.3\textwidth]{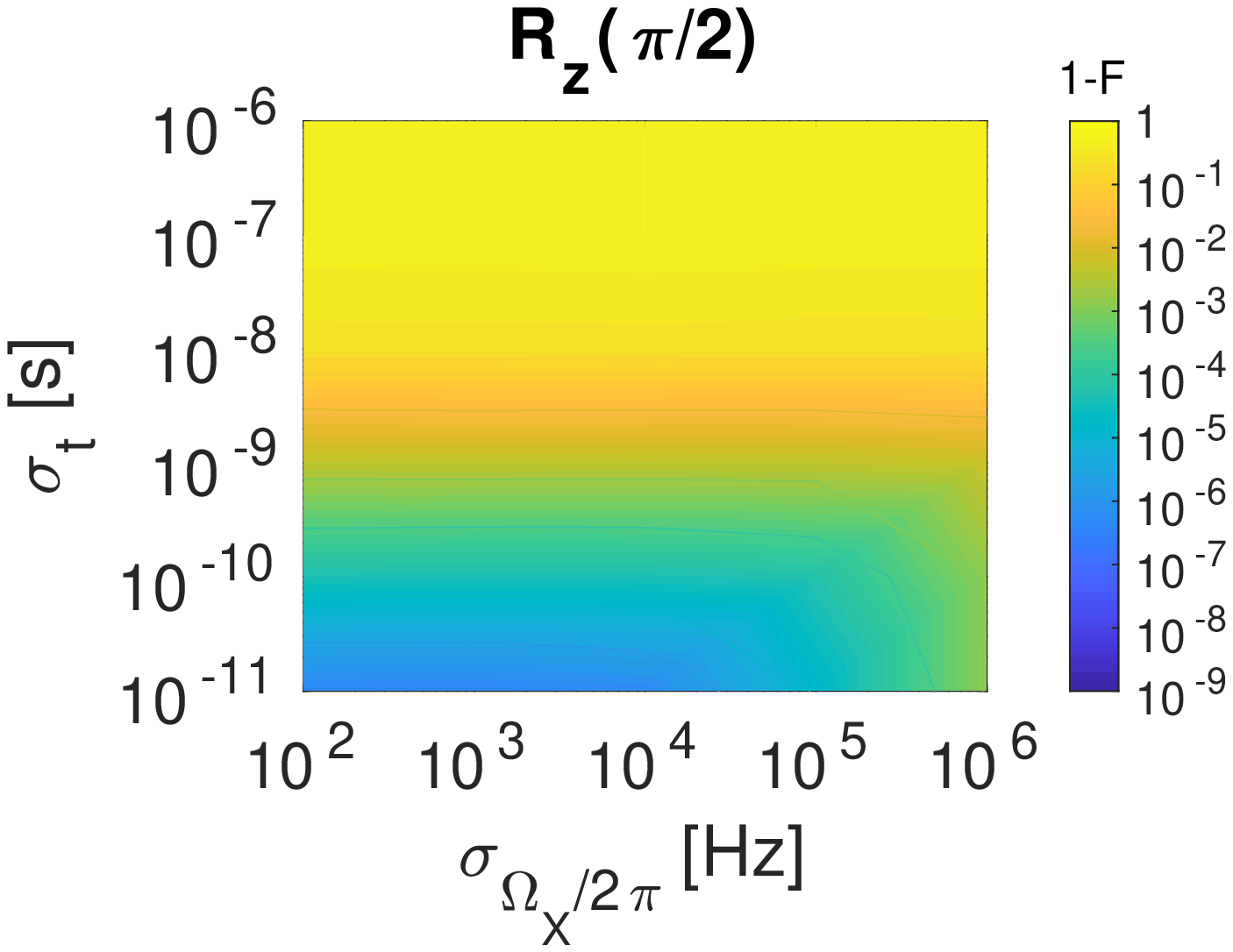}
		\includegraphics[width=0.3\textwidth]{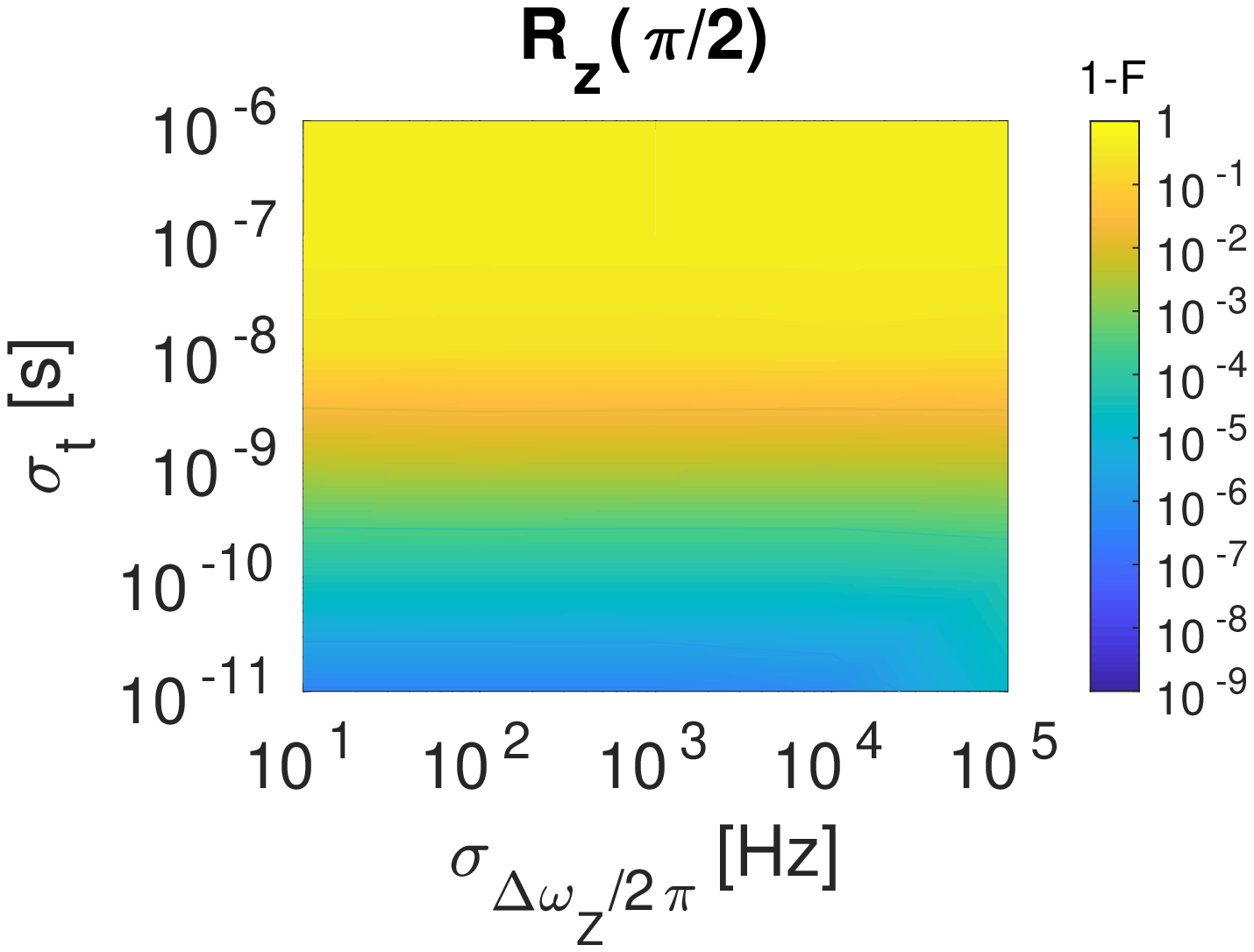}
		\includegraphics[width=0.3\textwidth]{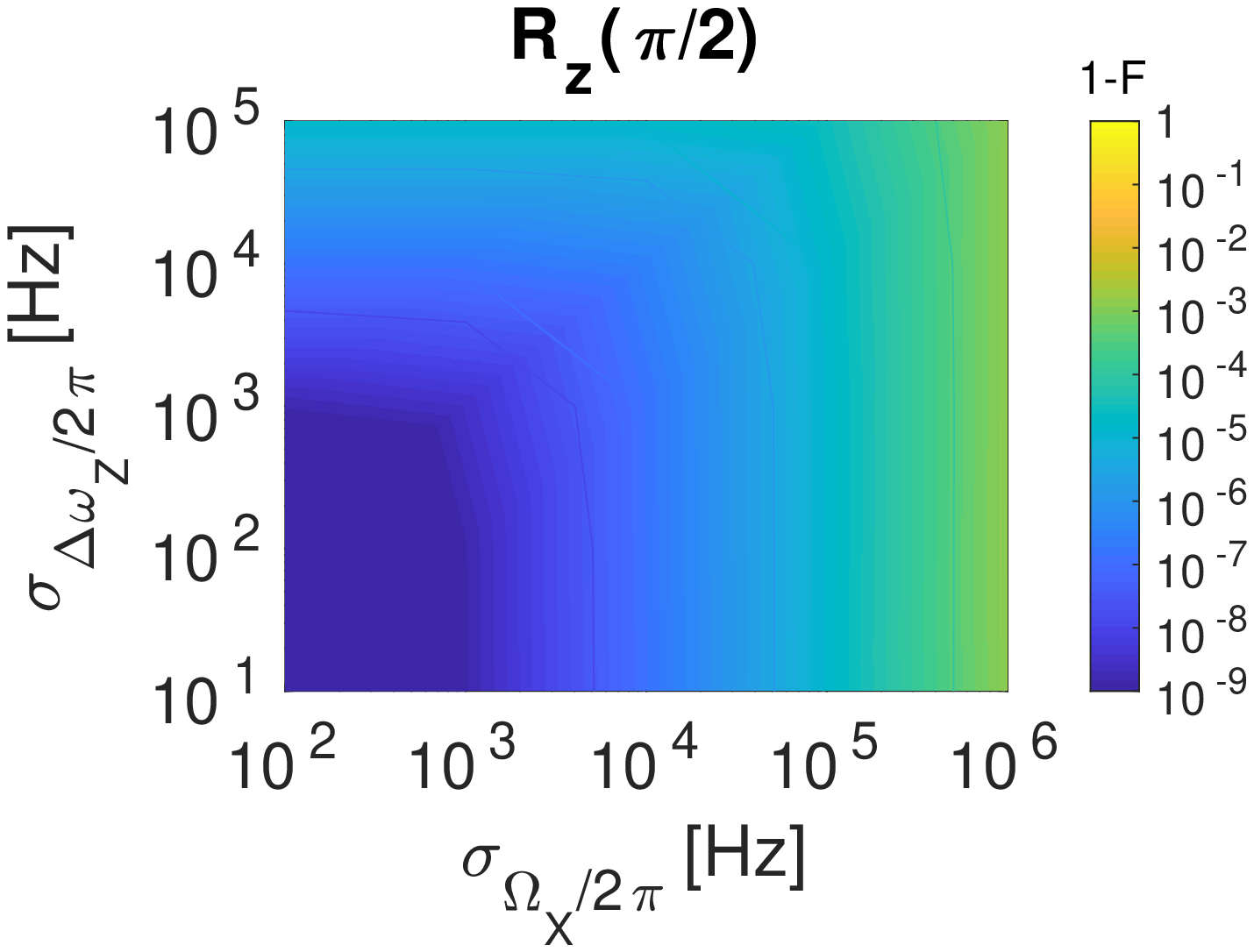}
	\end{center}
	\caption{Quantum dot spin qubit: a) 2D Plot showing $R_{x}(\pi/2)$ gate infidelity when $\sigma_{\Delta\omega_z/2\pi}$=0 (left), $\sigma_{\Omega_{x}/2\pi}$=0 (center) and $\sigma_t$=0 (right). b) Same as a) but for $R_{z}(\pi/2)$}\label{Fig:SQ} 
\end{figure}

The control variables of the SQ are the angular frequency $\Delta\omega_z$ and the angular frequency $\Omega_{x}$ that depends on the amplitude of the microwave used to obtain the ESR. The standard deviation ranges of the Gaussian distribution for the three random variables are set to: $\sigma_{\Delta\omega_z/2\pi}$ $\in$ [10, 10$^5$]  Hz, $\sigma_{\Omega_{x}/2\pi}$ $\in$ [10$^2$, 10$^6$] Hz and $\sigma_{t}$ $\in$ [10$^{-11}$, 10$^{-6}$ ] s.
Fig. \ref{Fig:SQ} shows plots of gate infidelities 1-F for $R_{x}(\pi/2)$ and $R_{z}(\pi/2)$ due to error sources. 
SQ has an infidelity that increases with respect to $\sigma_{t}$ for both operations. In particular lower infidelities can be achieved when $\sigma_t$=0, meaning that SQ is less sensitive to the error on the control signal amplitude ($\sigma_{\Delta\omega_z/2\pi}$ and $\sigma_{\Omega_{x}/2\pi}$) with respect to time interval error.

\subsection{Double quantum dot singlet-triplet qubit}
For the STQ the control variables are the exchange interaction $J$ and the additional Zeeman energy $\Delta E_z$. The ranges of standard deviations are set to: $\sigma_{\Delta E_z}$ $\in$ [10$^{-11}$, 10$^{-8}$] eV, $\sigma_{J}$ $\in$ [10$^{-10}$, 10$^{-7}$] eV and $\sigma_{t}$ $\in$ [10$^{-11}$ 10$^{-6}$] s. 
\begin{figure}[htbp]
	\begin{center}
		a)
		\includegraphics[width=0.3\textwidth]{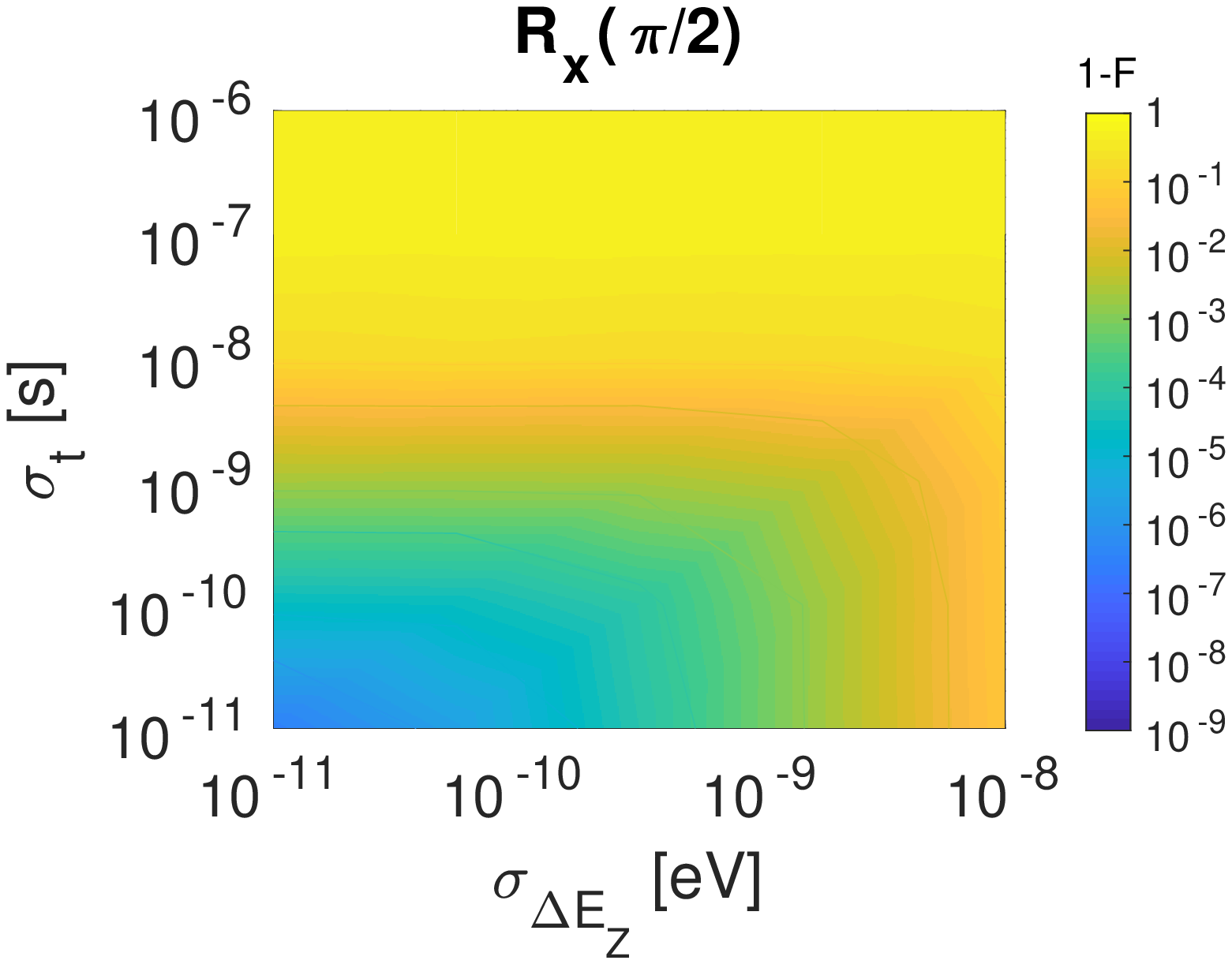}
		\includegraphics[width=0.3\textwidth]{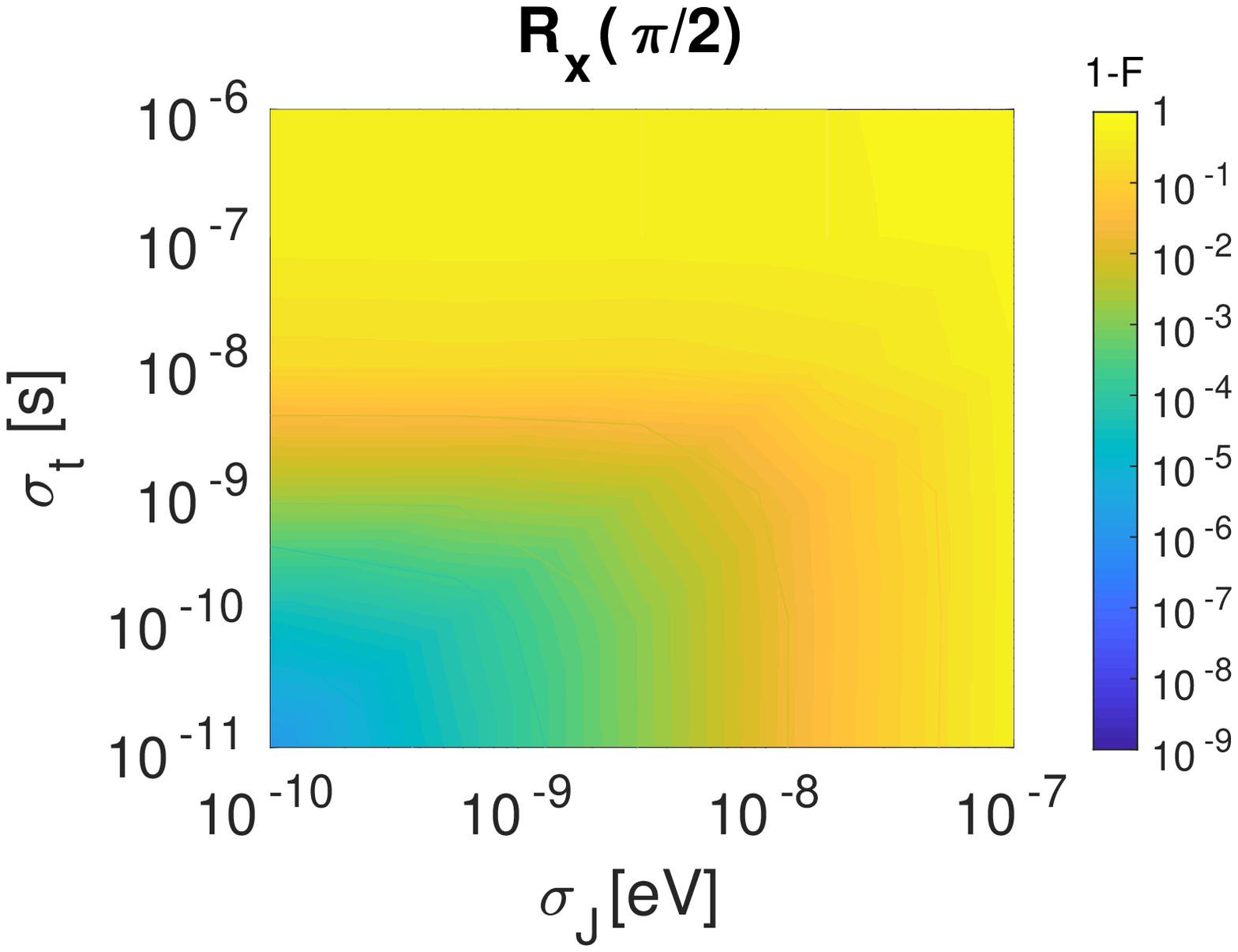}
		\includegraphics[width=0.3\textwidth]{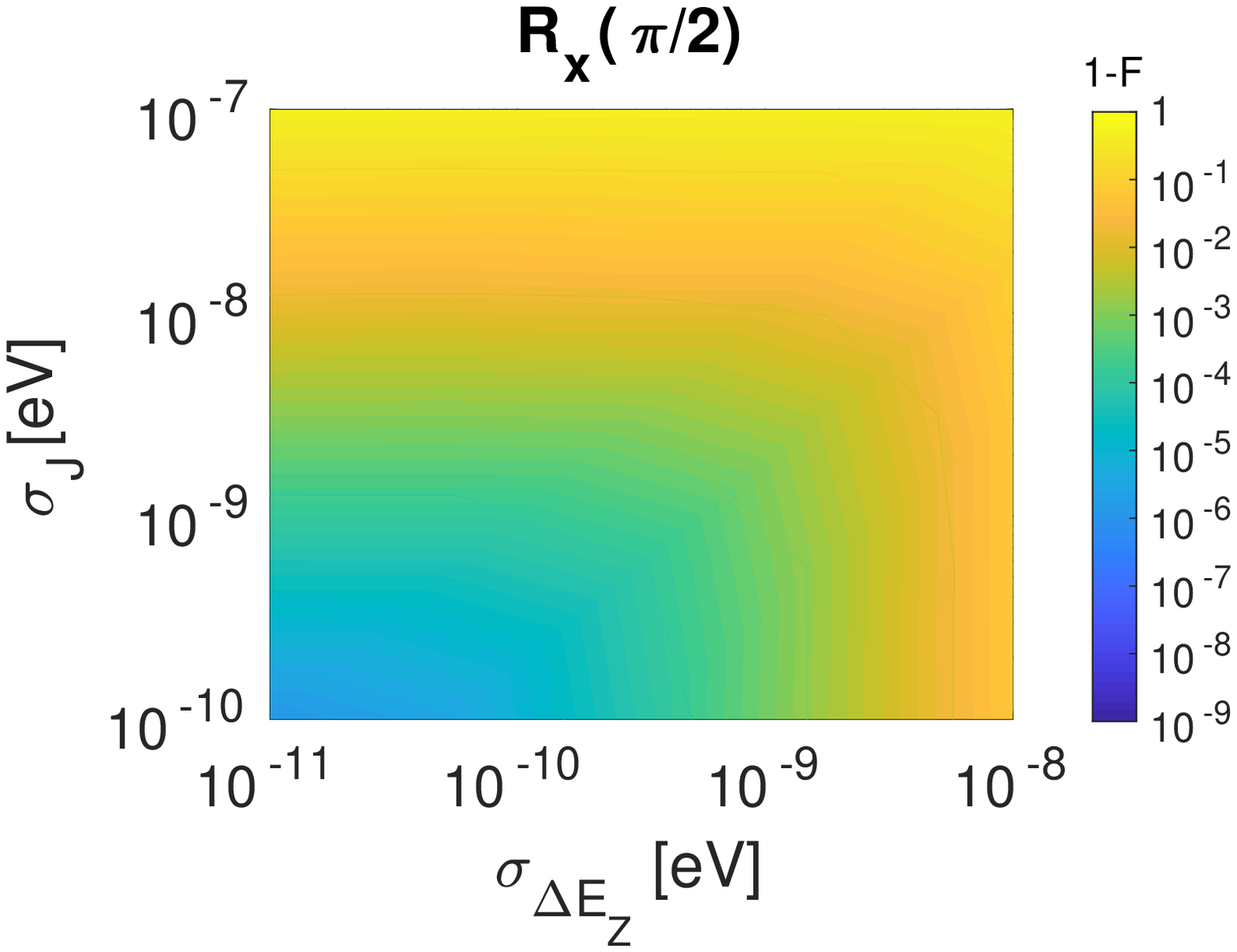}
	\end{center}
	\begin{center}
		b)
		\includegraphics[width=0.3\textwidth]{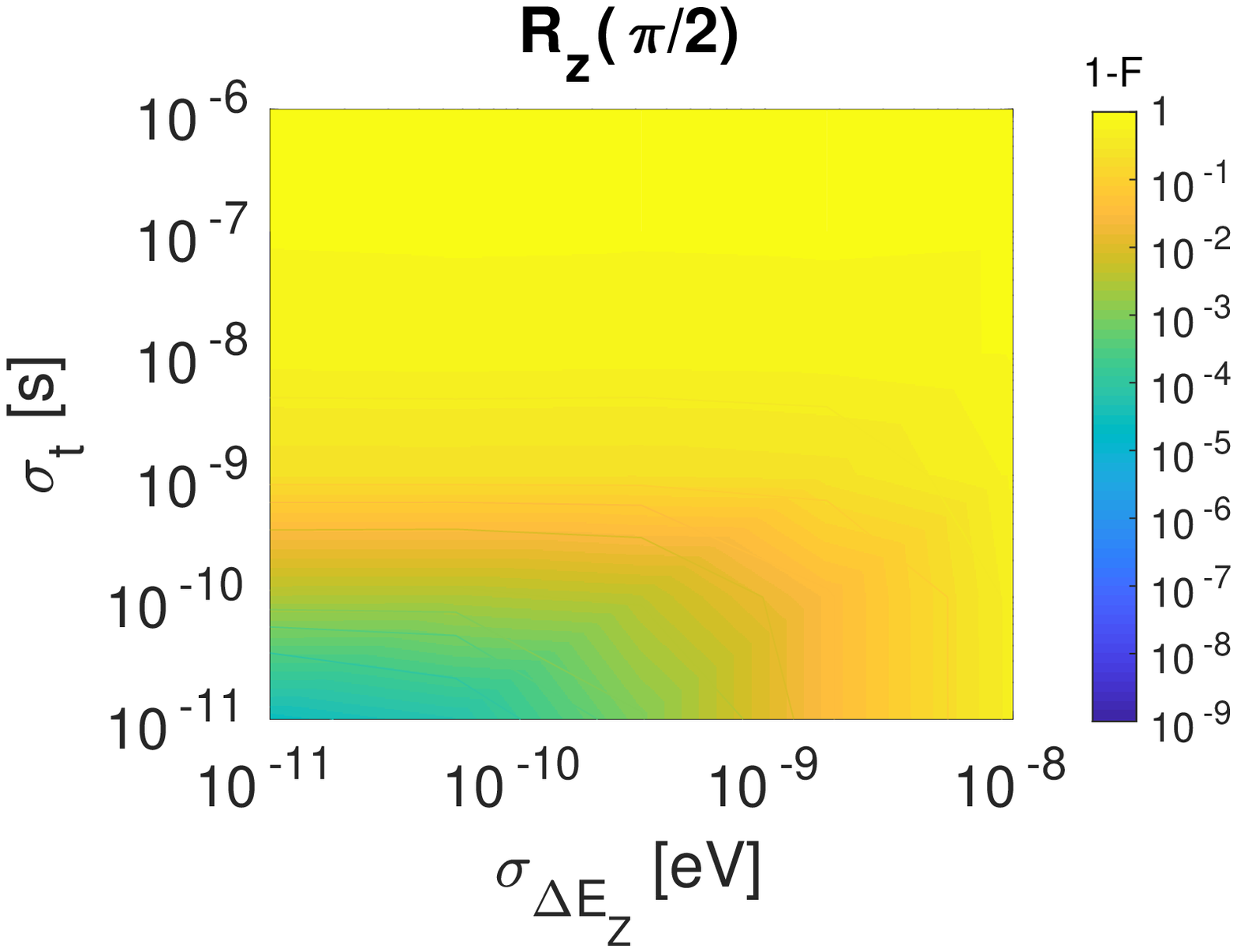}
		\includegraphics[width=0.3\textwidth]{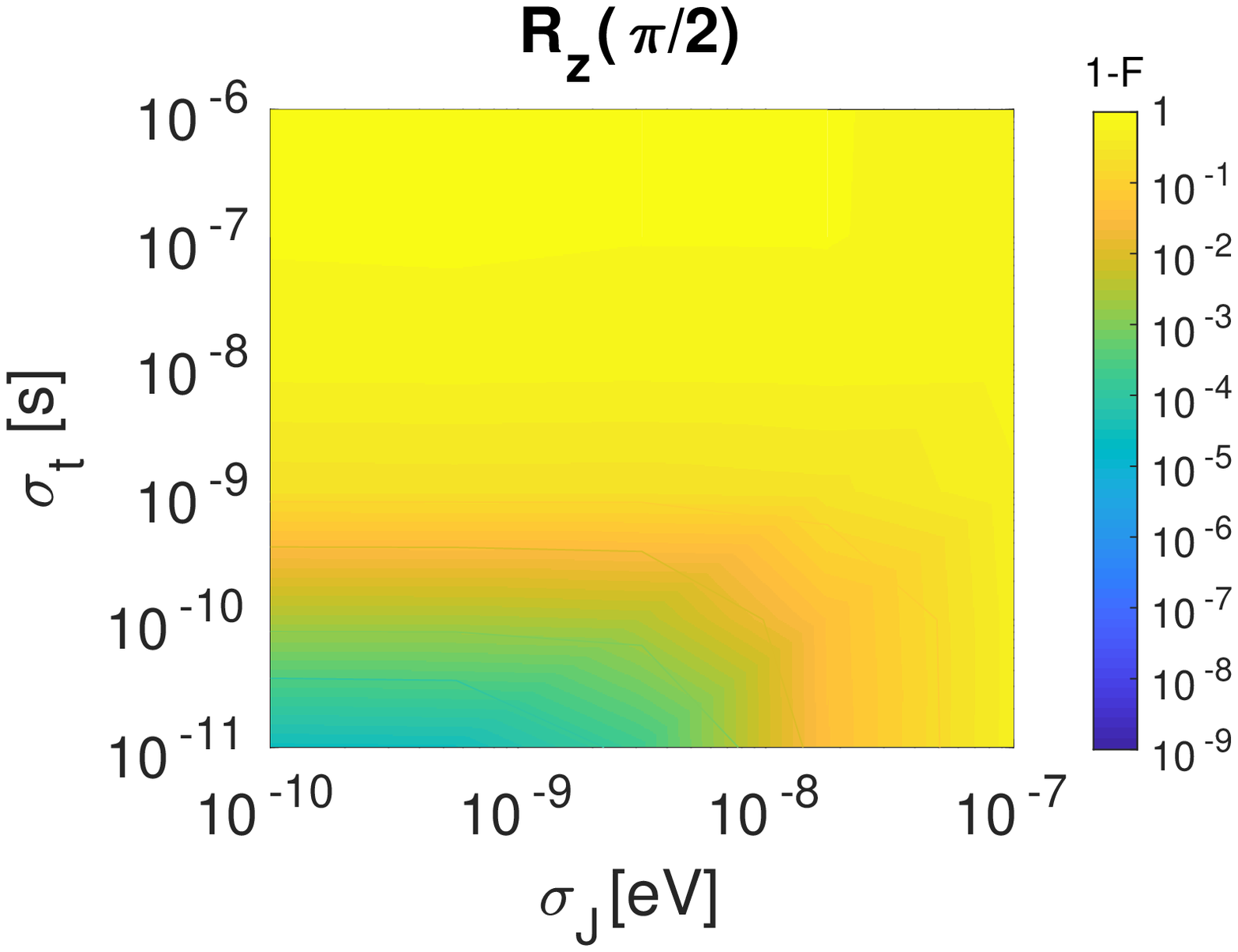}
		\includegraphics[width=0.3\textwidth]{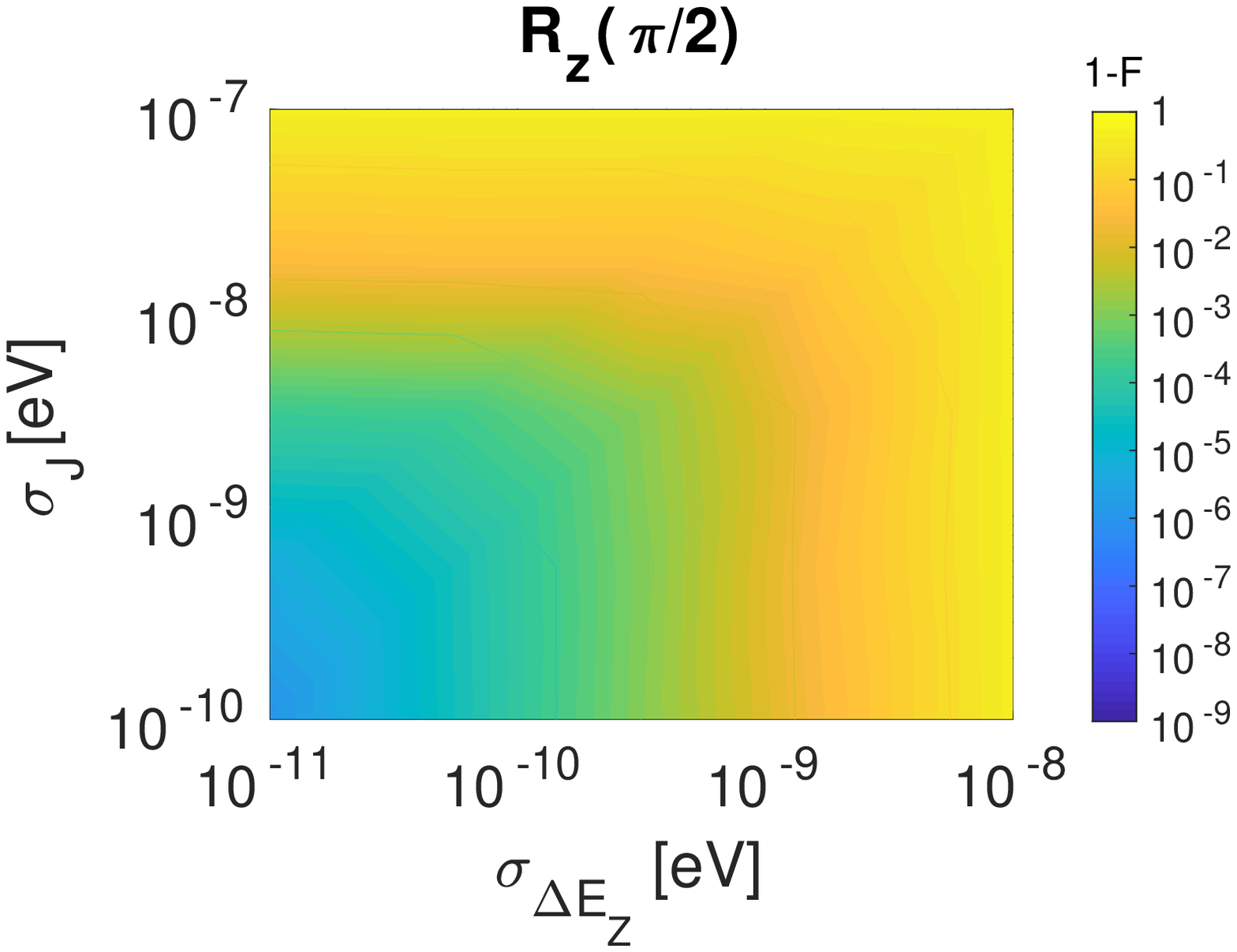}
	\end{center}
	\caption{Double quantum dot singlet-triplet qubit: a) 2D Plot showing $R_{x}(\pi/2)$ gate infidelity when $\sigma_J$=0 (left), $\sigma_{\Delta E_z}$=0 (center) and $\sigma_t$=0 (right). b) Same as a) but for $R_{z}(\pi/2)$.}\label{Fig:STQ} 
\end{figure}
As shown in Fig. \ref{Fig:STQ} STQ presents low infidelities for both gates in all the range of error considered. Both gate infidelities show a strong increment due to high $\sigma_{t}$. $R_{z}(\pi/2)$ gate has an infidelity that depend mainly on $\sigma_{J}$.   

\subsection{Double quantum dot hybrid qubit}
The HQ is controlled by a sequence of pulses of $J$ so the maximum exchange interaction $J^{max}$ is the control variable. The ranges of standard deviations set to calculate the infidelities for $R_{x}(\pi/2)$ and $R_{z}(\pi/2)$ are: $\sigma_{J}$ $\in$ [10$^{-10}$, 10$^{-6}$] eV and $\sigma_{t}$ $\in$ [10$^{-11}$, 10$^{-6}$] s. 
\begin{figure}[h!]
	\begin{center}
		a)
		\includegraphics[width=0.45\textwidth]{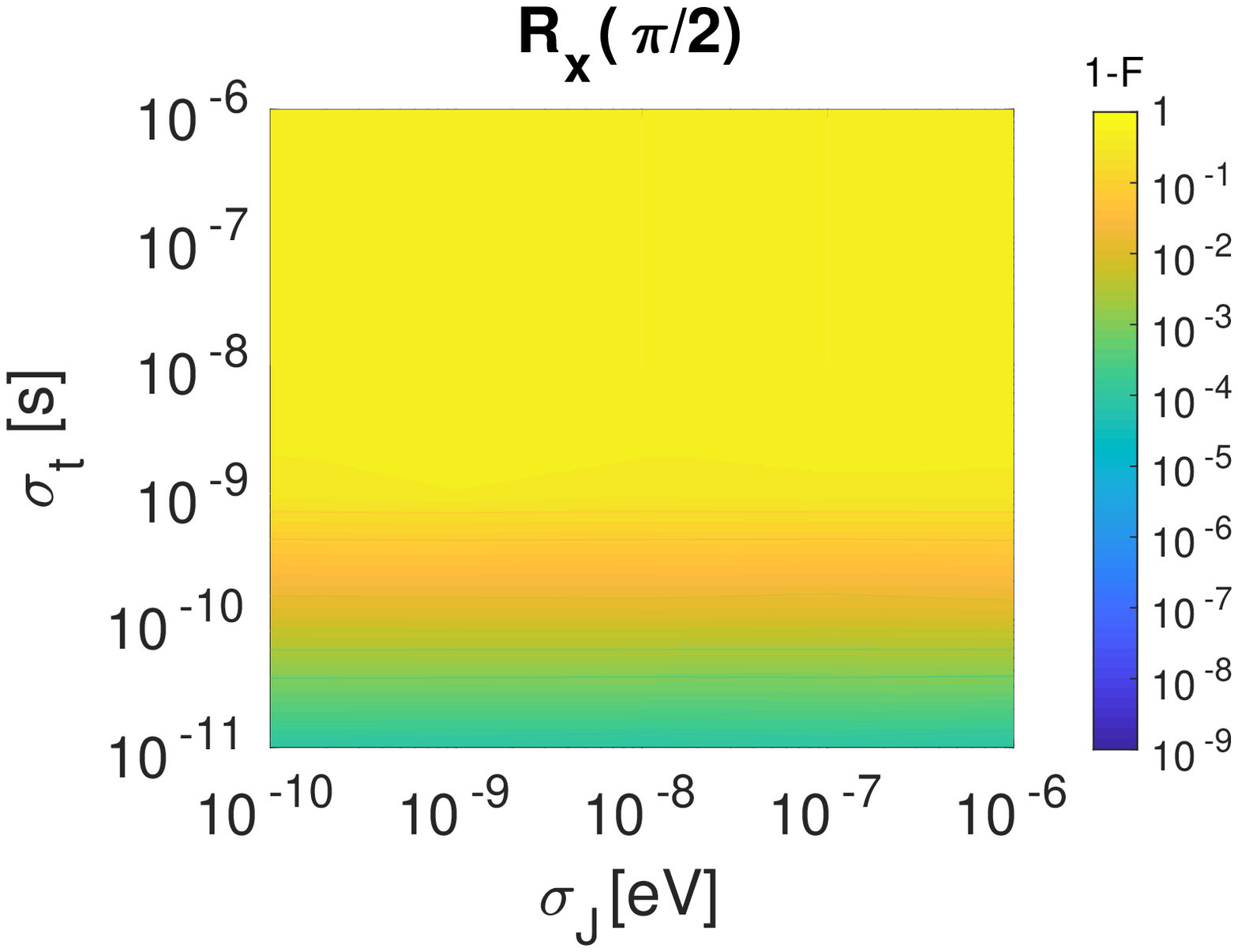}
		b)	
		\includegraphics[width=0.45\textwidth]{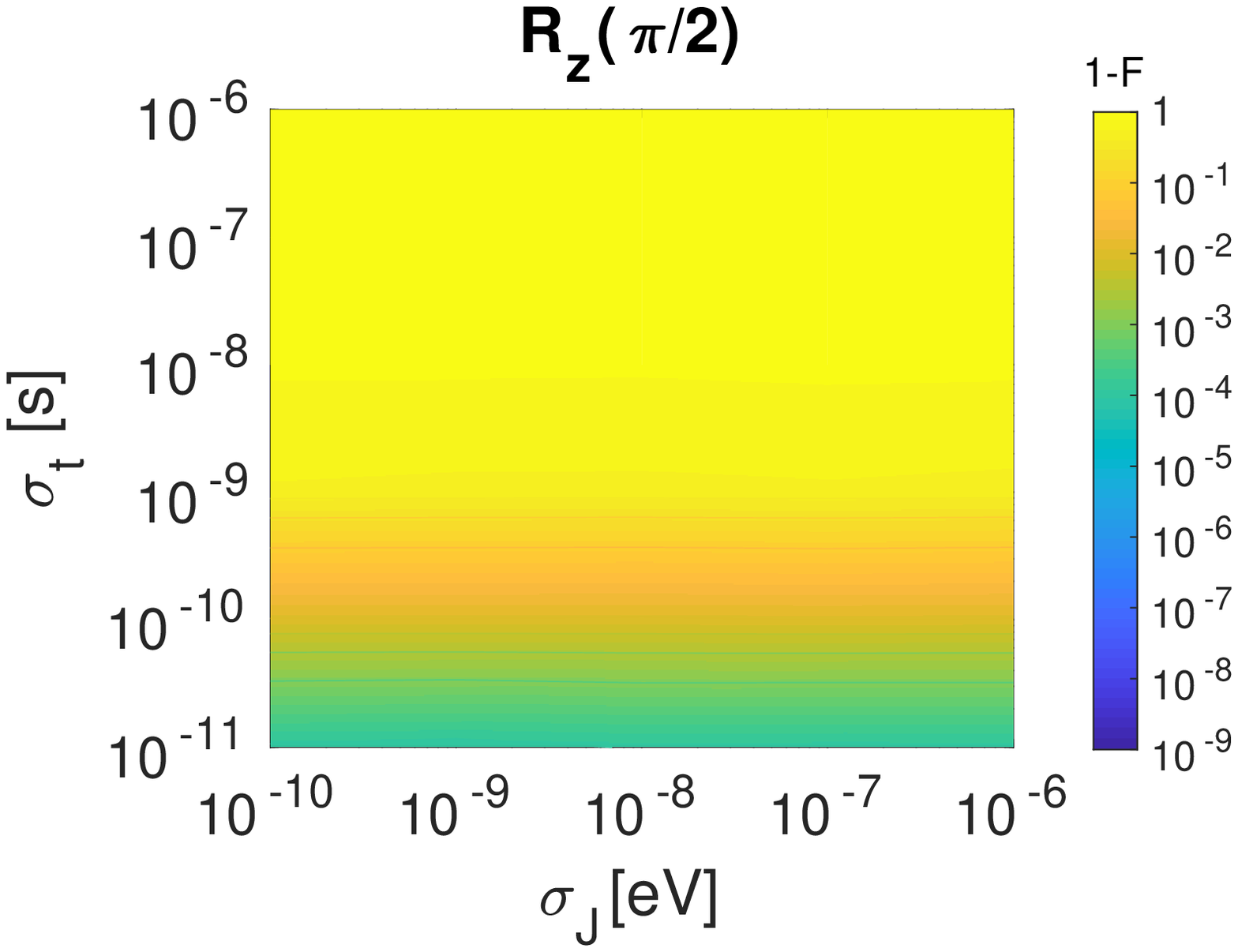}
	\end{center}
	\caption{Double quantum dot hybrid qubit: a) 2D Plot showing $R_{x}(\pi/2)$ gate infidelity due to $\sigma_{J}$ and $\sigma_{t}$.  b) Same as a) but for $R_{z}(\pi/2)$.}\label{Fig:HQ} 	
\end{figure}
In Fig. \ref{Fig:HQ} HQ shows a strong dependence of infidelity on $\sigma_{t}$ for both gates whereas no evident effect can be highlighted in the range of $\sigma_{J}$ considered.  

\subsection{Donor qubit}
For the DQ the control variables are the angular frequency $\Omega_{x}$ that depends on the amplitude of the microwave and the frequency detuning $\Delta\omega_{12}=\omega_{12}-\omega$, that is the difference between the microwave and the voltage-modulated hyperfine interaction $A$ angular frequencies. The standard deviation of the distribution of the corresponding random variables are set to: $\sigma_{\Delta \omega_{12}/2\pi}$ $\in$ [10, 10$^5$]  Hz, $\sigma_{ \Omega_{x}/2\pi}$ $\in$ [10, 10$^5$] Hz and $\sigma_{t}$ $\in$ [10$^{-11}$, 10$^{-6}$ ] s.
\begin{figure}[h!]
	\begin{center}
		a)
		\includegraphics[width=0.3\textwidth]{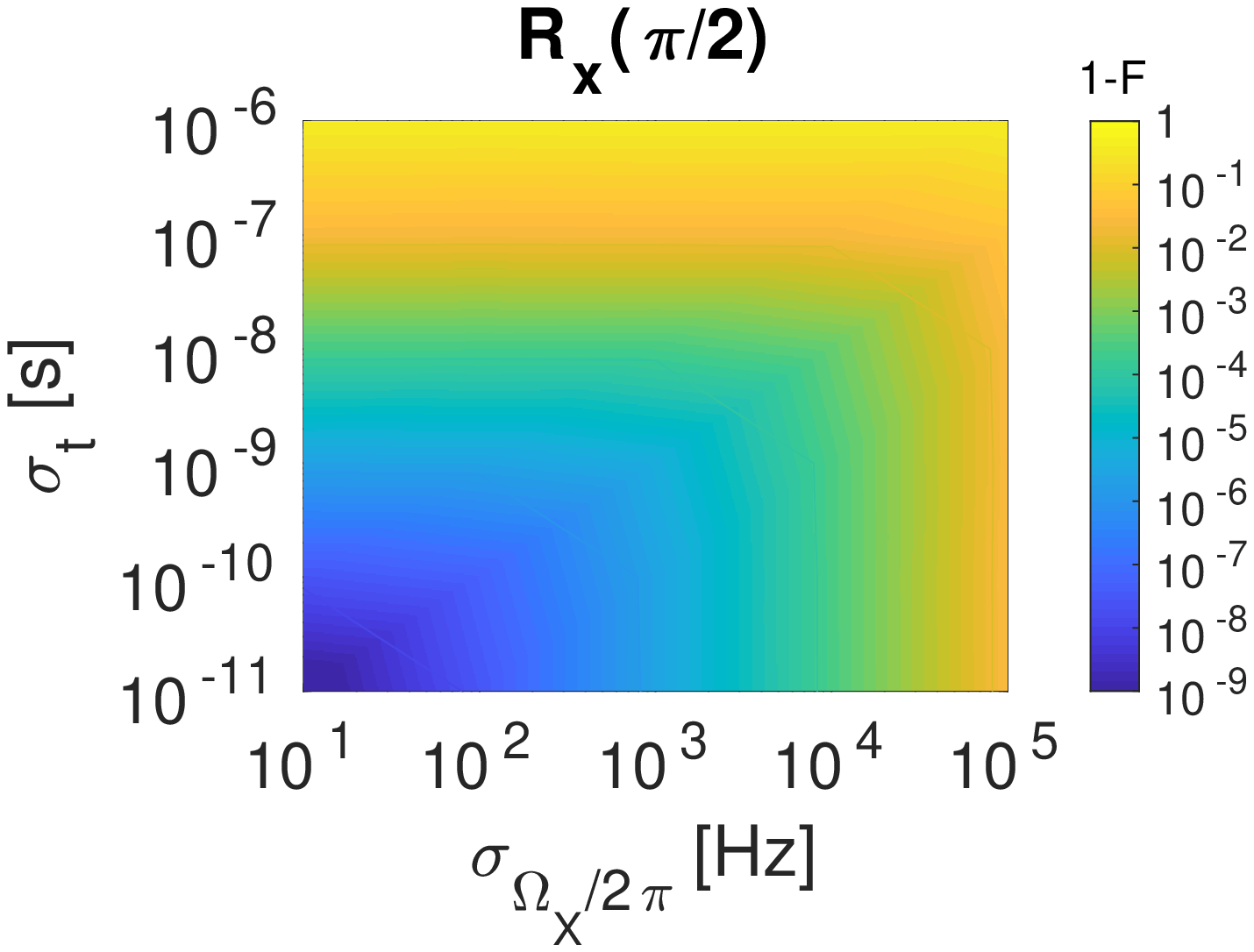}
		\includegraphics[width=0.3\textwidth]{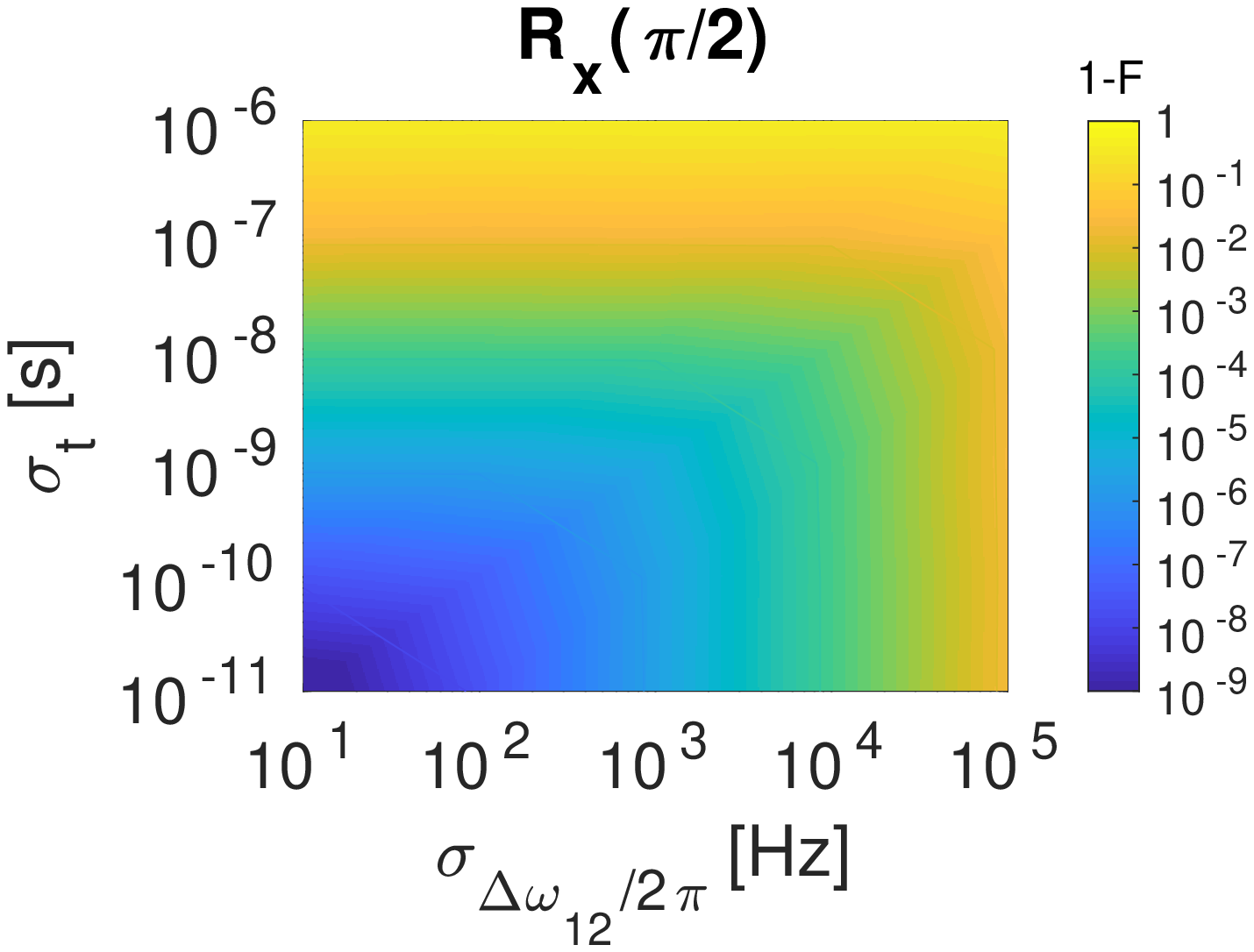}
		\includegraphics[width=0.3\textwidth]{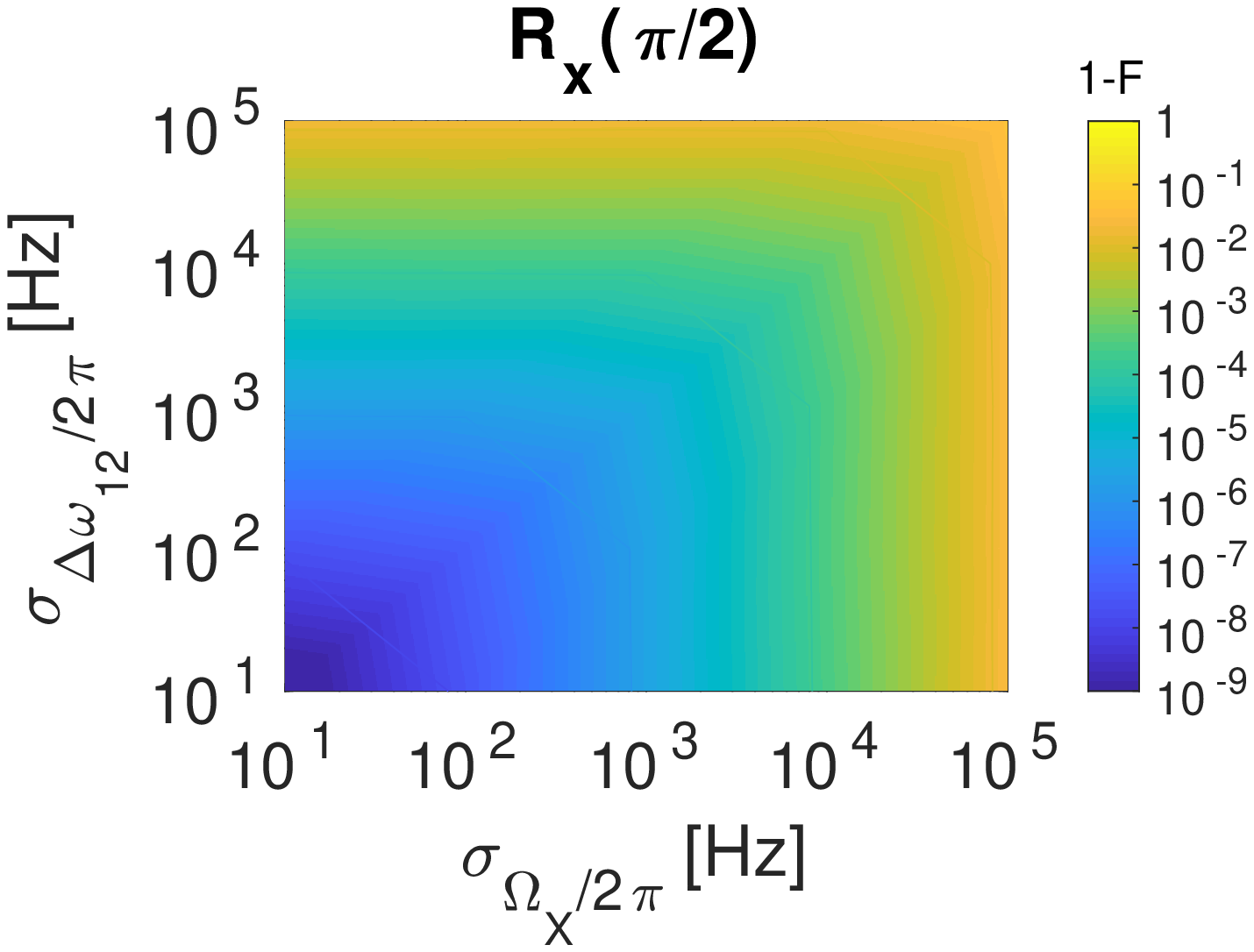}
	\end{center}
	\begin{center}
		b)
		\includegraphics[width=0.3\textwidth]{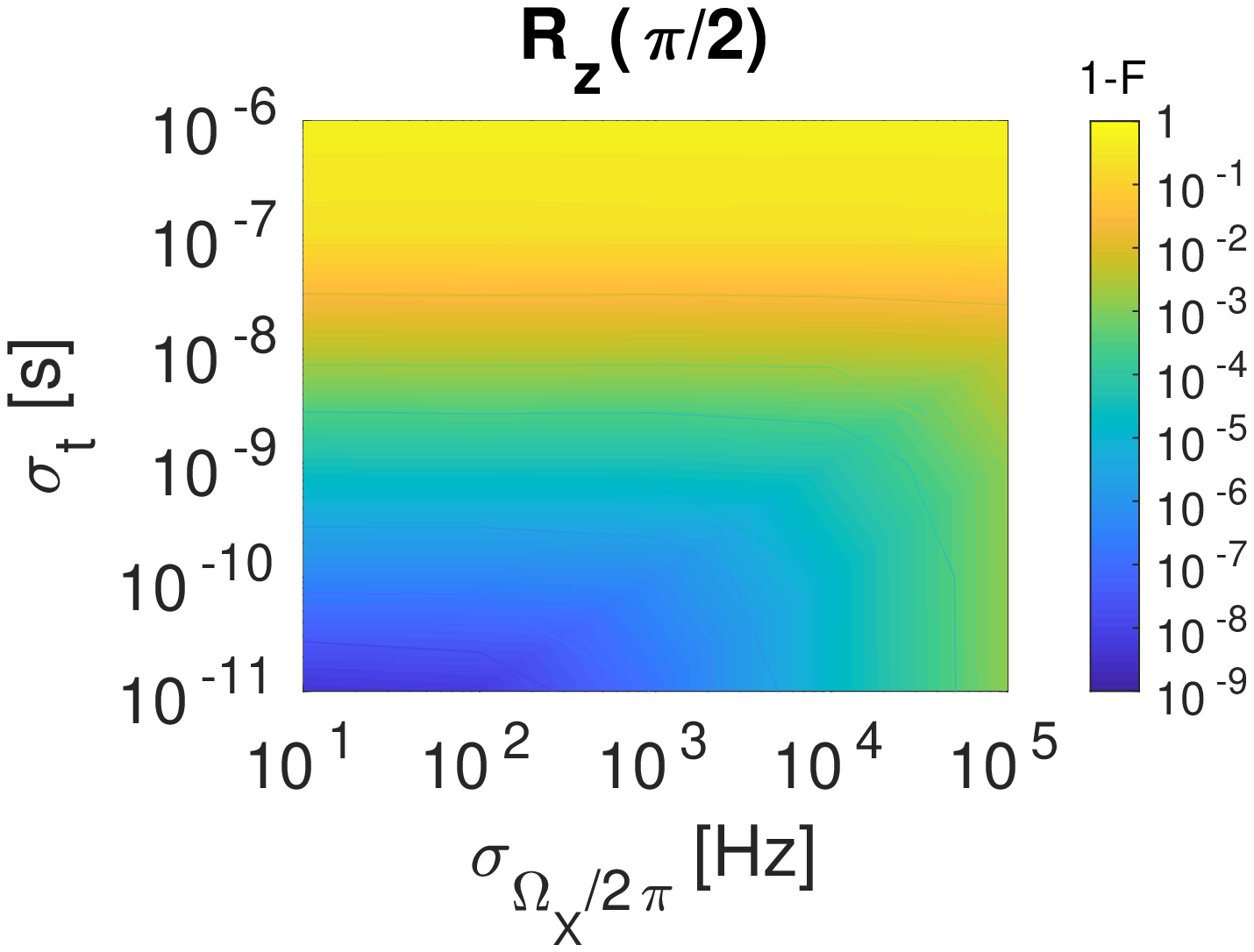}
		\includegraphics[width=0.3\textwidth]{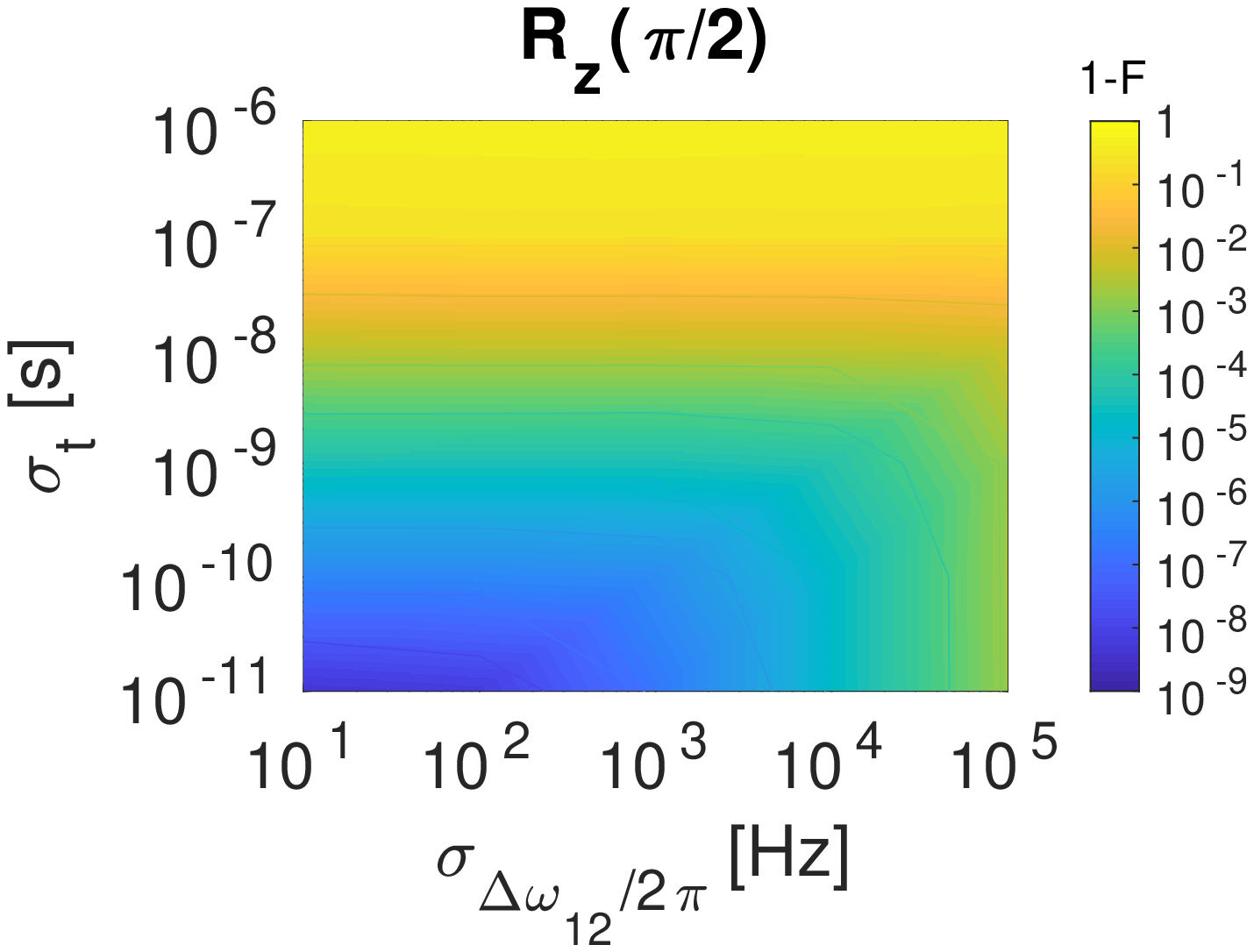}
		\includegraphics[width=0.3\textwidth]{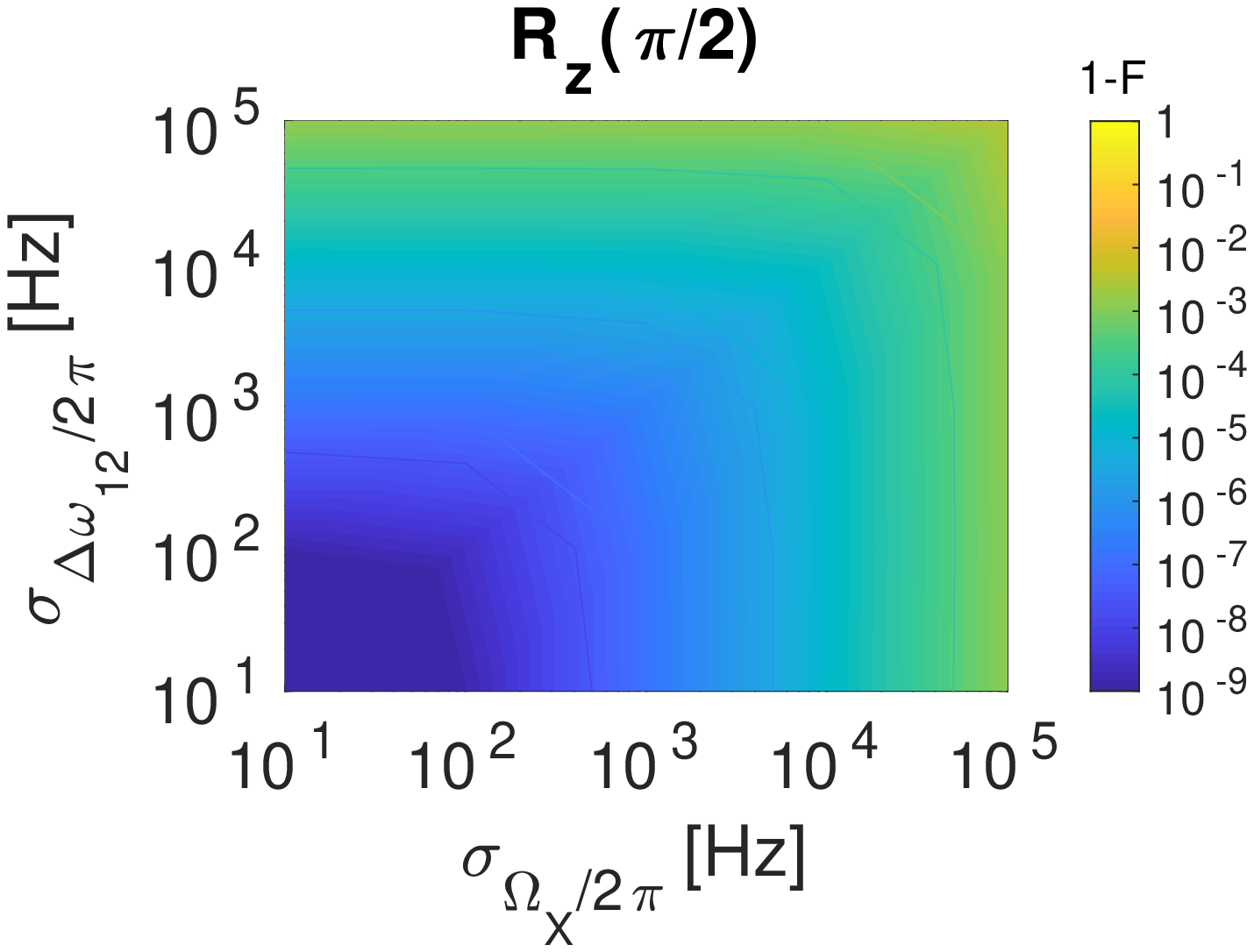}
	\end{center}
	\caption{Donor qubit: a) 2D Plot showing $R_{x}(\pi/2)$ gate infidelity due to $\sigma_{\Omega_{x}/2\pi}$ and $\sigma_{t}$ with $\sigma_{\Delta \omega_{12}/2\pi}$ (left), $\sigma_{\Delta \omega_{12}/2\pi}$ and $\sigma_{t}$ with $\sigma_{\Omega_{x}/2\pi}$=0 (center), $\sigma_{\Omega_{x}/2\pi}$ and $\sigma_{\Delta \omega/2\pi}$ with $\sigma_{t}$=0 (right).  b) Same as a) but for $R_{z}(\pi/2)$.}\label{Fig:DQ} 
\end{figure}
DQ has a quite low infidelity in the $\sigma_{t}$ range studied for $R_{x}(\pi/2)$ gate and $R_{z}(\pi/2)$ gate as reported in Fig. \ref{Fig:DQ}. This is directly connected to the step times reported in Tab. \ref{Tab:Allsequences} for both gate operations where long pulses are used.

\subsection{Quantum dot spin-donor qubit}
The exchange interaction $J$ and the voltage modulated hyperfine interaction $A$ are the control variables of the SDQ with associated random variables with standard deviation in ranges equal to: $\sigma_{A}$ $\in$ [10$^{-11}$, 10$^{-7}$] eV, $\sigma_{J}$ $\in$ [10$^{-11}$, 10$^{-7}$] eV and $\sigma_{t}$ $\in$ [10$^{-11}$, 10$^{-6}$] s. 
\begin{figure}[h!]
	\begin{center}
		a)
		\includegraphics[width=0.3\textwidth]{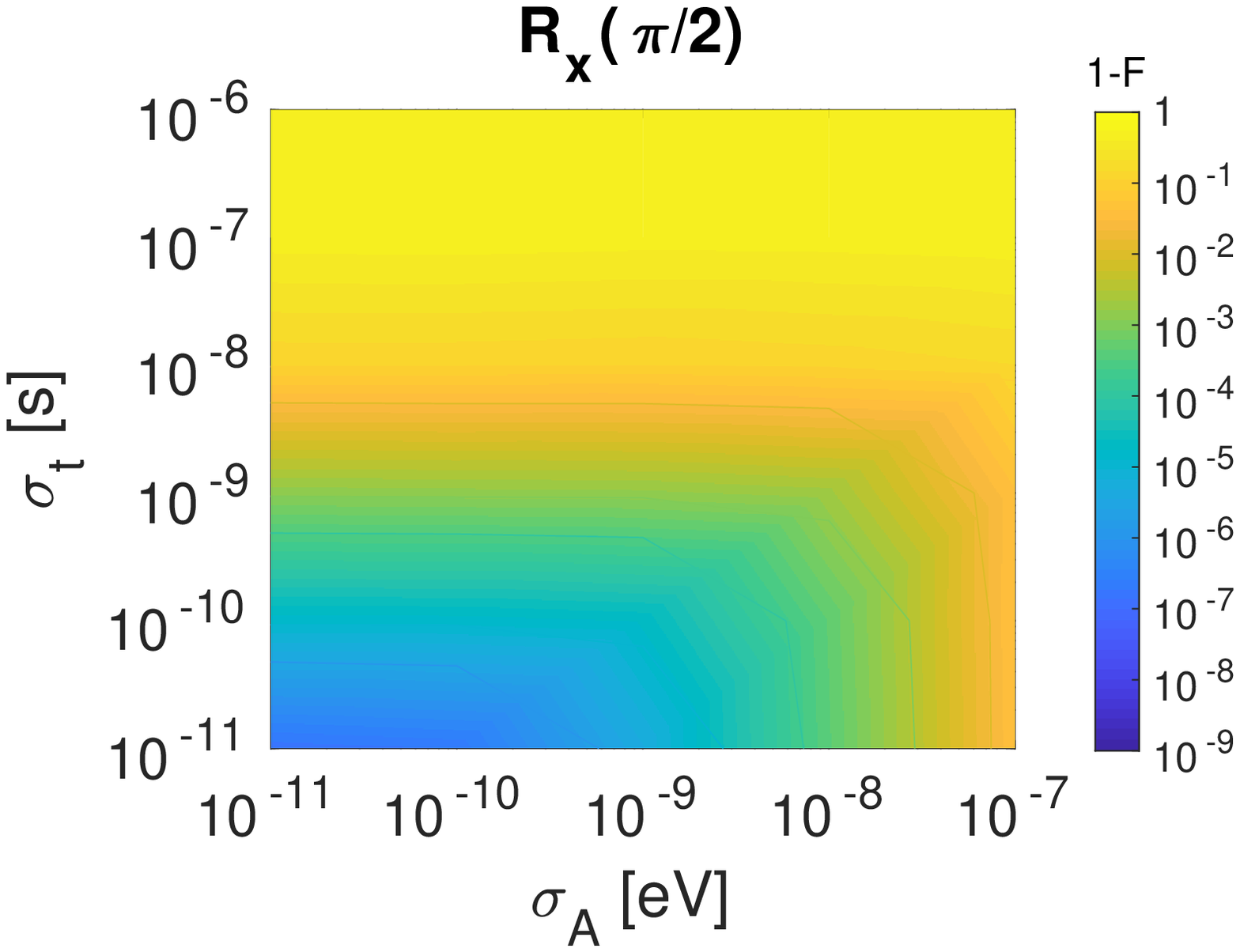}
		\includegraphics[width=0.3\textwidth]{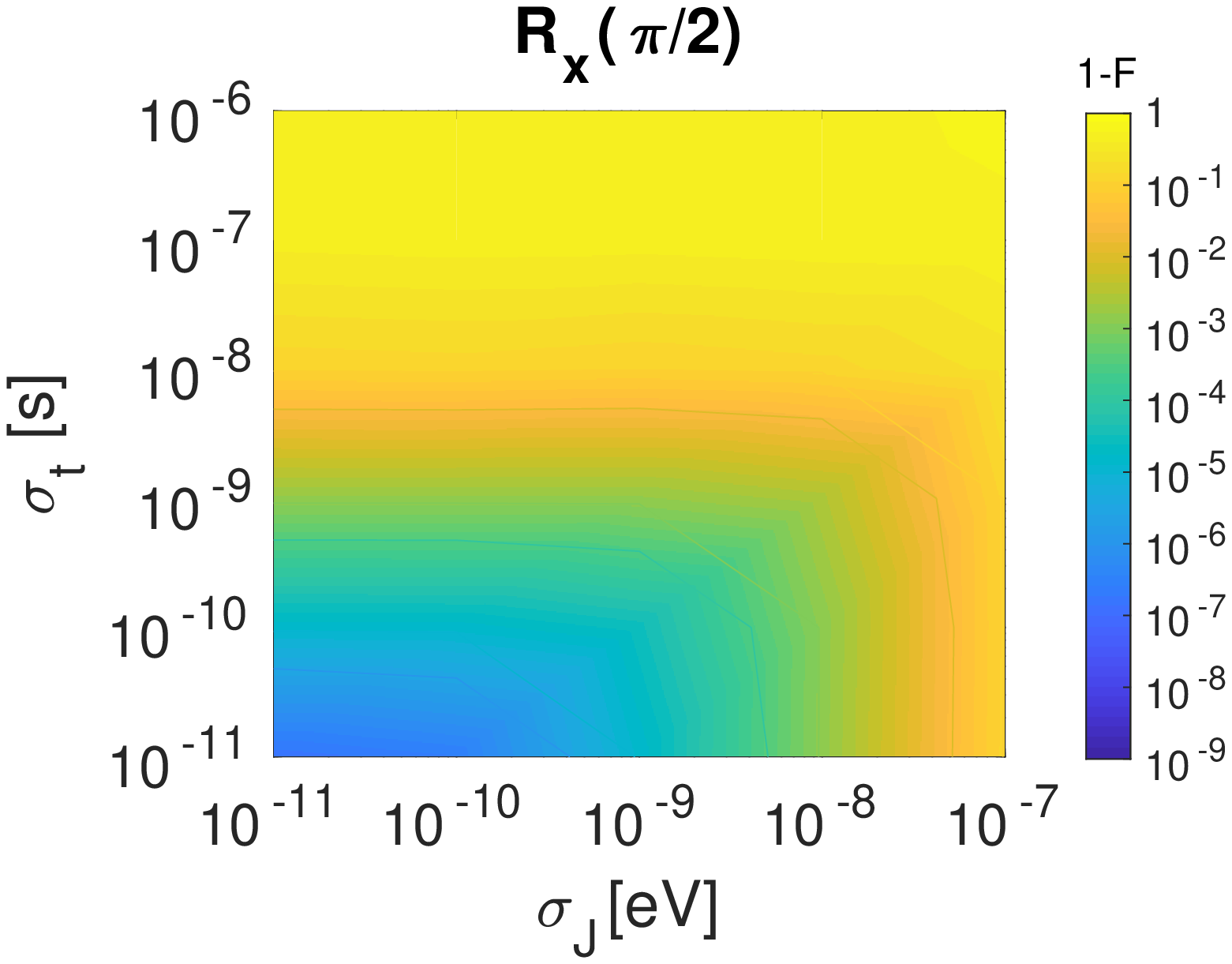}
		\includegraphics[width=0.3\textwidth]{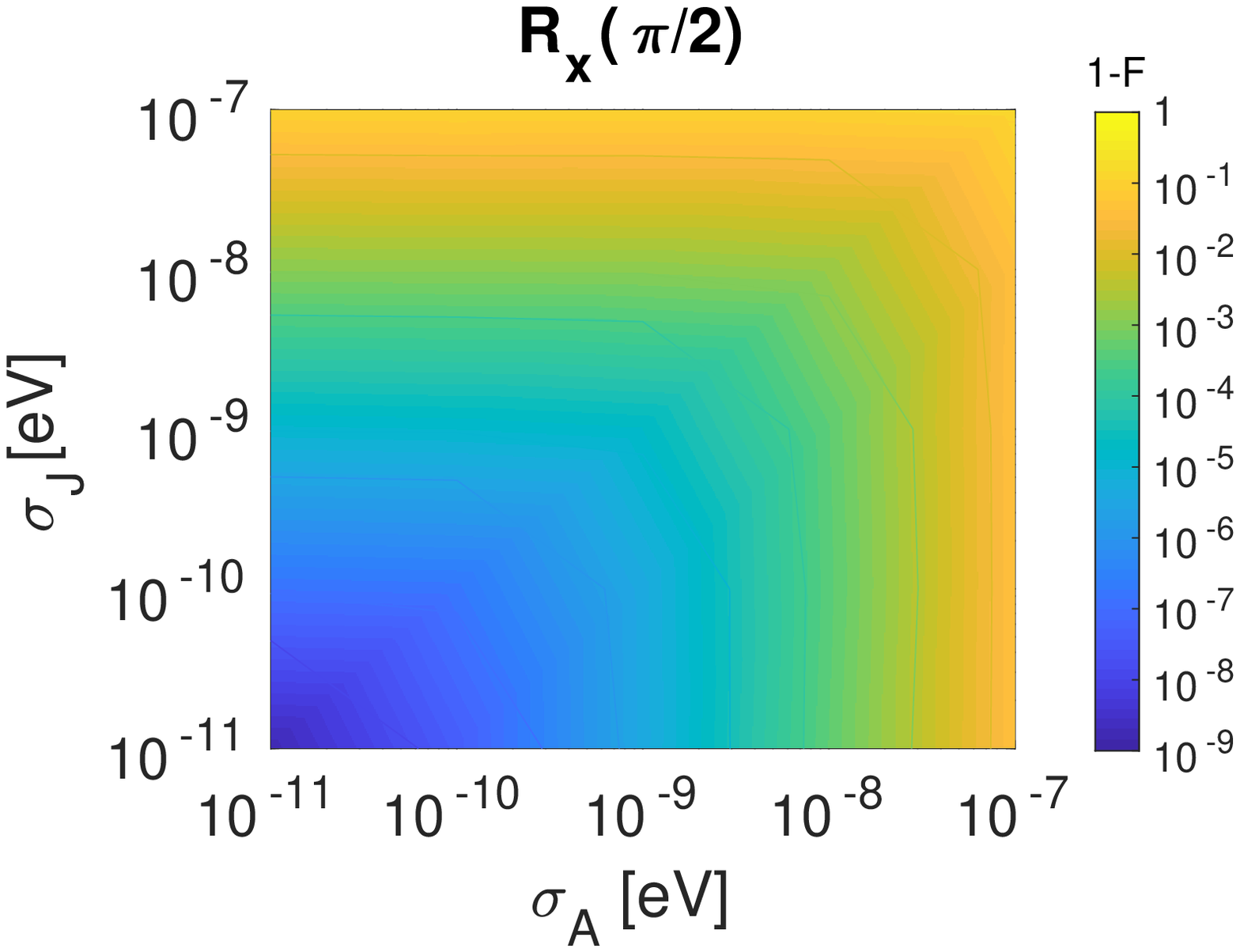}
	\end{center}
	\begin{center}
		b)
		\includegraphics[width=0.3\textwidth]{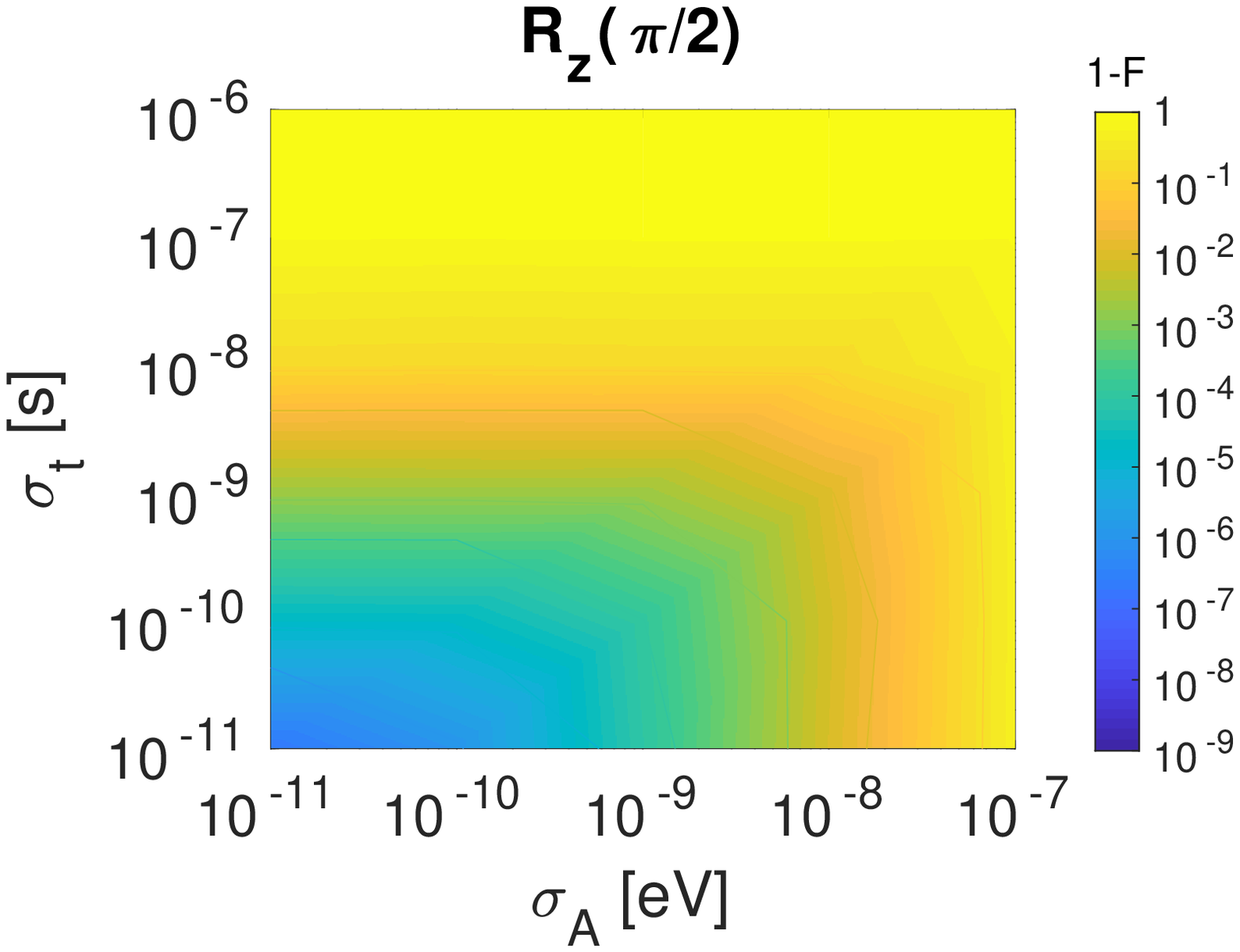}
		\includegraphics[width=0.3\textwidth]{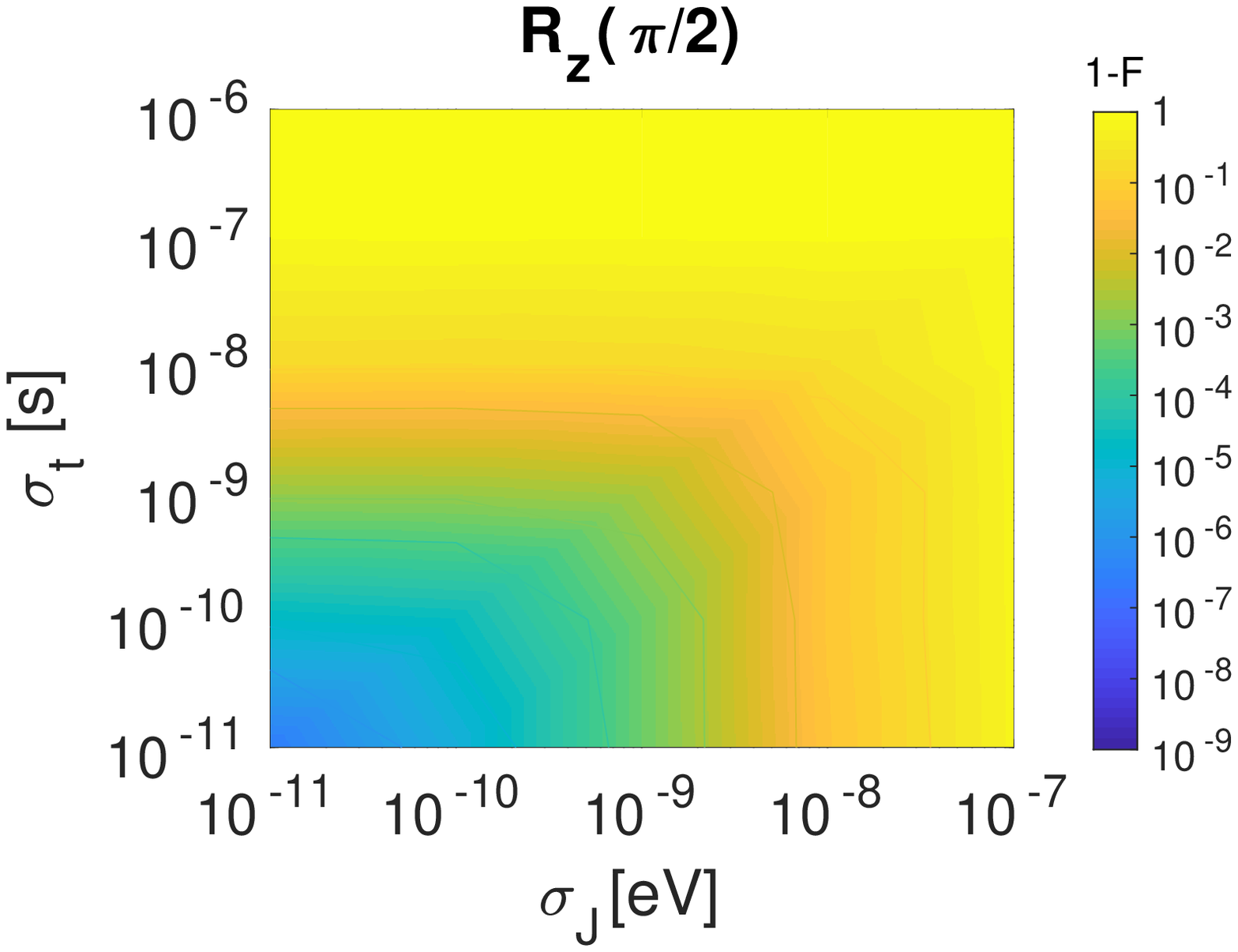}
		\includegraphics[width=0.3\textwidth]{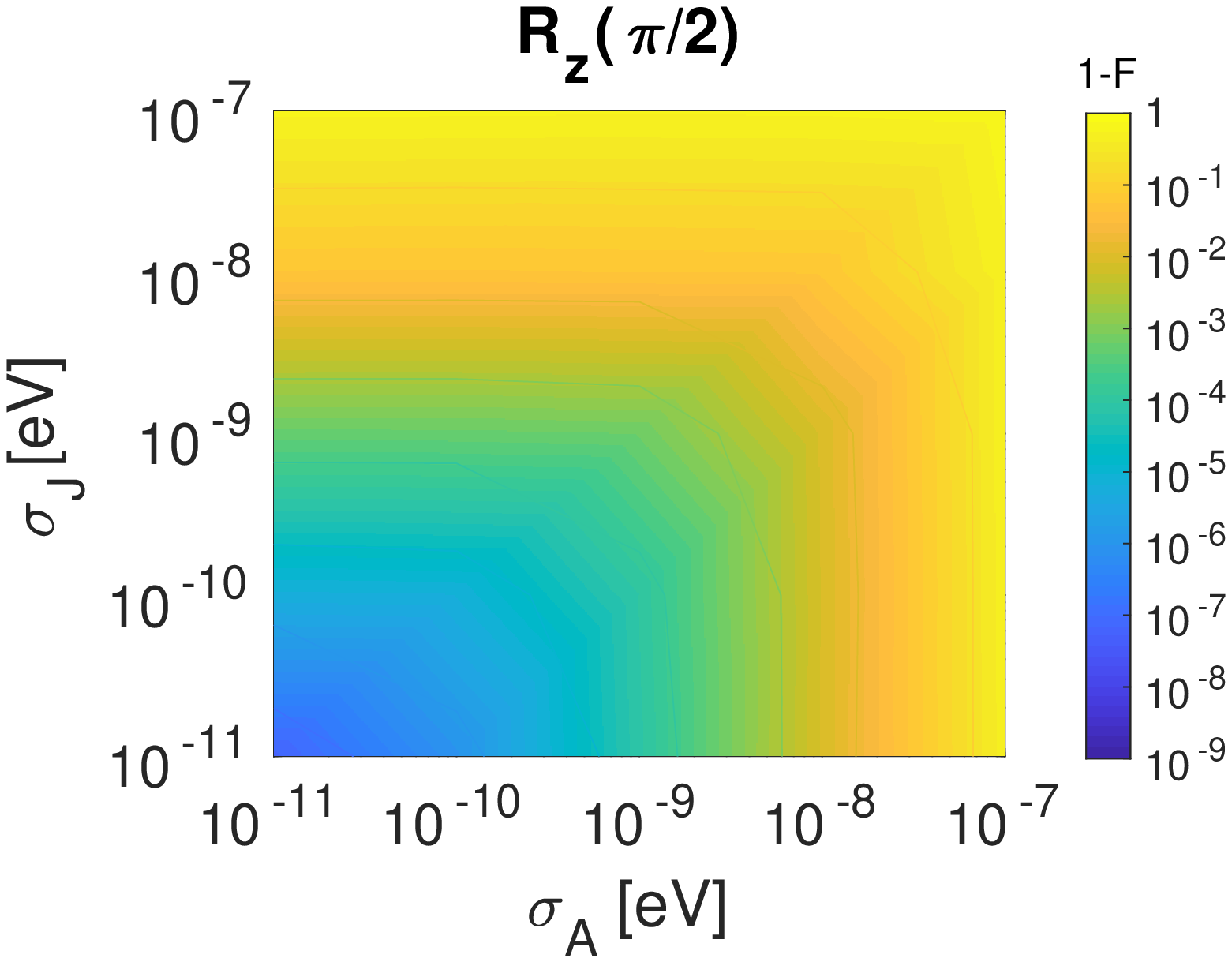}
	\end{center}
	\caption{Quantum dot spin-donor qubit: a) 2D Plot showing $R_{x}(\pi/2)$ gate infidelity when $\sigma_{J}$=0 (left), $\sigma_{A}$=0 (center) and $\sigma_{t}$=0 (right).  b) Same as a) but for  $R_{z}(\pi/2)$.}\label{Fig:SDQ} 
\end{figure}
SDQ shows low infidelities in the considered error ranges for both gates (Fig. \ref{Fig:SDQ}). $R_{x}(\pi/2)$ shows a lower infidelity due to $\sigma_{J}$ than $R_{z}(\pi/2)$ gate.

\section{Comparison of gate fidelities among qubit types}
In this section, a comparison on gate fidelities between all the qubit types due to the TIE is presented. To this purpose, control error standard deviations on the amplitudes of the control parameters are set to the values reported in Tab. \ref{Tab:ErrorRanges}. 
\begin{table}
	\caption{\label{Tab:ErrorRanges} Control error standard deviations for the five qubit types. The parameter values are collected from the literature.}
	\begin{indented}
		\item[]\begin{tabular}{@{}lll}
			\br
			Qubit  & Error on control variables  &  \\ 
			\mr
		 	SQ & $\sigma_{\Delta\omega_z/2\pi}$ = 20 Hz \cite{DatasheetMWgenerator} & $\sigma_{\Omega_x/2\pi}$=0.25 MHz \cite{Kawakami-2014} \\ 
			STQ & $\sigma_{\Delta E_z}$=4 neV \cite{XianWu-2014} & $\sigma_{J}$=1 neV \cite{XianWu-2014} \\ 
			HQ & $\sigma_{J}$=1 neV \cite{Thorgrimsson-2017}&   \\ 
			DQ & $\sigma_{\Delta \omega_{12}/2\pi}$=100 Hz \cite{Laucht-2015} & $\sigma_{\Omega_x/2\pi}$=25 kHz \\ 
			SDQ & $\sigma_{J}$=4 neV \cite{Harvey-2017} & $\sigma_{A}$=2.5 neV \cite{Harvey-2017} \\ 
			\br
		\end{tabular}
	\end{indented}
\end{table}

\begin{figure}[htbp]
	\begin{center}
a) 	\includegraphics[width=0.45\textwidth]{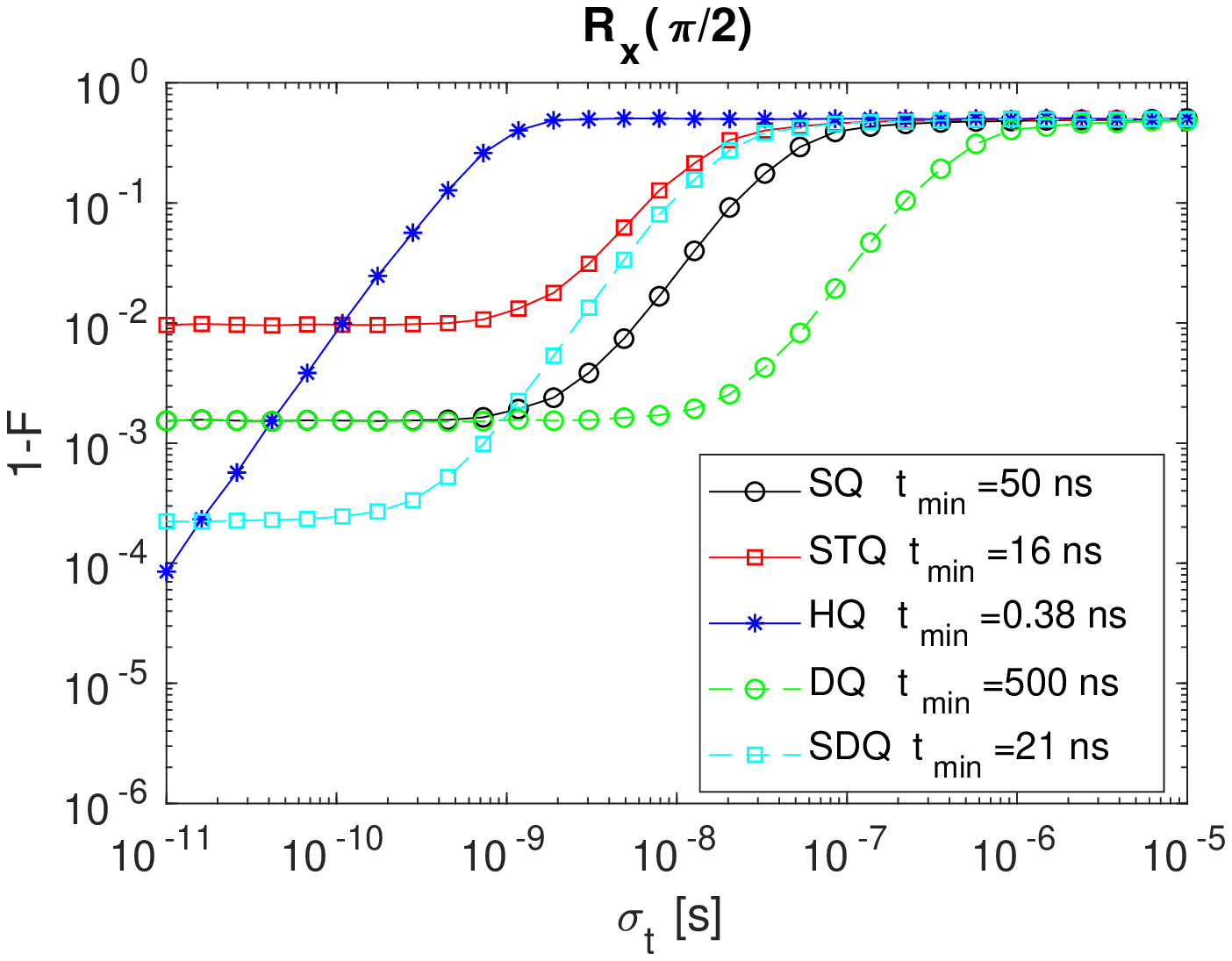}
b) 	\includegraphics[width=0.45\textwidth]{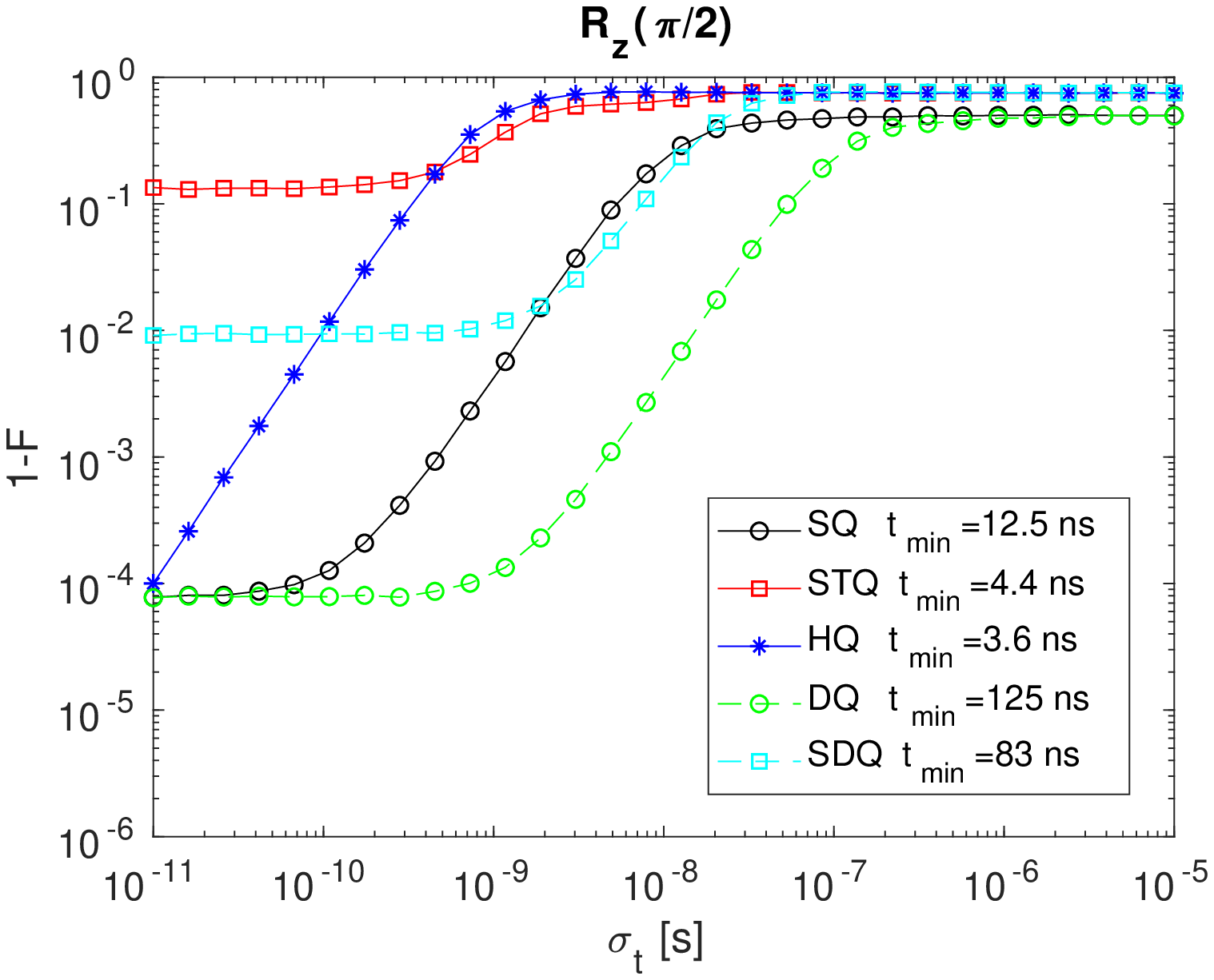}
	\end{center}
	\caption{a) Comparison of $R_{x}(\pi/2)$ gate infidelities among all qubit types as a function of the standard deviation $\sigma_{t}$. In the legend the smallest time of sequence step $t_{min}$ for each qubit type is also reported. b) Same as a) but for $R_{z}(\pi/2)$ gate.}\label{Fig:Fcomparison_Rot} 
\end{figure}

Fig. \ref{Fig:Fcomparison_Rot} shows a gate infidelities comparison for $R_{x}(\pi/2)$ and $R_{z}(\pi/2)$. 

For both gates, all the qubit types generally show decreasing infidelities when $\sigma_t$ is reduced. The roll off of each curve is roughly observed for $\sigma_t$ close to the shortest step time $t_{min}$ of the gate sequence for the corresponding qubit type. Note that HQ is the most sensitive qubit to TIE whereas DQ is the most robust to such kind of error for both operations. But such robustness of the DQ is achieved by imposing slower gates than those of other qubits (see the columns of step and total time in Tab. \ref{Tab:Allsequences}). In other words, given a qubit type, there is a trade-off between the sequence step time and the robustness of the gate fidelity to TIE.
SQ, STQ, DQ and SDQ present a saturated behavior when $\sigma_t$ is reduced meaning that the TIE is no more the fidelity limiter in that range. On the contrary, the infidelities of HQ do not saturate even for small $\sigma_t$.

The study presented depends on the physical parameters used in the calculations as reported in Tables \ref{Tab:Allsequences} and \ref{Tab:ErrorRanges}, that are at the state of the art from experiments. The analysis is obviously susceptible of variation if control and error amplitudes are varied. 

Future studies could include extended comparisons among gate infidelity of qubits manipulated with pulse sequences properly designed to mitigate the effects of control errors. Such sequences could be obtained by exploiting optimal control algorithms, such as the GRAPE algorithm \cite{Khaneja-2005}.
 
\section{Conclusions}
In this paper we reported a comparative study of five spin qubit types realized through the spin of electrons in electrostatically defined quantum dots, and through the spin of electrons of impurity atoms in the semiconducting host (donors). The implementations studied are: the quantum dot spin qubit, the double quantum dot singlet-triplet qubit, the double quantum dot hybrid qubit, the donor qubit and the quantum dot spin-donor qubit. For each qubit type, we derived analytical time sequences that realize single qubit rotations along the principal axis of the Bloch sphere. Then, by using a Gaussian noise model, the effects on the gate fidelity of the errors disturbing the control parameters are estimated. A comparison of the gate fidelities of all the qubit implementations due to the time interval error are presented using a realistic set of values for the error parameters of amplitude controls taken from the literature. By using those parameter values at the state of the art we can conclude that the hybrid qubit infidelity is very sensitive to TIE because qubit rotations are achieved with very fast sequences. Conversely, the infidelity of the donor qubit due to TIE is not dominant till very large time errors, at a cost of quite slow gate operations. This study offers a general platform to investigate gate fidelities in different spin qubit implementations and gives an important instrument for future works together with the development of experimental realizations.     

\ack
This work was supported by the European Union's Horizon 2020 research and innovation program under grant agreement No 688539 MOS-QUITO. The authors would thank M. Belli for fruitful discussions.

\vspace{10pt}

\section*{References}
\bibliographystyle{iopart-num}
\bibliography{Ref}

\providecommand{\newblock}{}
\begin{thebibliography}{10}
\expandafter\ifx\csname url\endcsname\relax
  \def\url#1{{\tt #1}}\fi
\expandafter\ifx\csname urlprefix\endcsname\relax\def\urlprefix{URL }\fi
\providecommand{\eprint}[2][]{\url{#2}}

\bibitem{Shulman-2012}
Shulman M, Dial O, Harvey S, Bluhm H, Umansky V and Yacoby A 2012 {\em
  Science\/} {\bf 336} 202

\bibitem{Veldhorst-2014}
Veldhorst M, Hwang J~C~C, Yang C~H, Leenstra A~W, de~Ronde B, Dehollain J~P, T
  J, Hudson F~E, Itoh K~M, Morello A and Dzurak A~S 2014 {\em Nature
  Nanotechnology\/} {\bf 9} 981--985

\bibitem{Pla-2012}
Pla J, Tan K, Dehollain J, Lim W, Morton J, Jamieson D, Dzurak A and Morello A
  2012 {\em Nature\/} {\bf 489} 541

\bibitem{Maune-2012}
Maune B~M, Borselli M~G, Huang B, Ladd T~D, Deelman P~W, Holabird K~S, Kiselev
  A~A, Alvarado-Rodriguez I, Ross R~S, Schmitz A~E, Sokolich M, Watson C~A,
  Gyure M~F and Hunter A~T 2012 {\em Nature\/} {\bf 481} 344--347

\bibitem{Bluhm-2011}
Bluhm H, Foletti S, Neder I, Rudner M, Mahalu D, Umansky V and Yacoby A 2011
  {\em Nature Physics\/} {\bf 7} 109

\bibitem{Tyryshkin-2012}
Tyryshkin A, Tojo S, Morton J, Riemann H, Abrosimov N, Becker P, Pohl H~J,
  Schenkel T, Thewalt M, Itoh K and Lyon S 2012 {\em Nature Material\/} {\bf
  11} 143

\bibitem{RuiLi-2012}
Li R, Hu X and You J 2012 {\em Physical Review B\/} {\bf 86} 205306

\bibitem{Coish-2005}
Coish W and Loss D 2005 {\em Physical Review B\/} {\bf 72} 125337

\bibitem{Shen-2000}
Shen S~Q and Wang Z 2000 {\em Physical Review B\/} {\bf 61} 9532

\bibitem{Morton-2011}
Morton J~J~L, McCamey D~R, Eriksson M~A and Lyon S~A 2011 {\em Nature\/} {\bf
  479} 345--353

\bibitem{Kawakami-2014}
Kawakami E, Scarlino P, Ward D~R, Braakman F~R, Savage D~E, Lagally M~G,
  Friesen M, Coppersmith S~N, Eriksson M~A and Vandersypen L~M~K 2014 {\em
  Nature Nanotechnology\/} {\bf 9} 666--670

\bibitem{Klymenko-2015}
Klymenko M~V, Rogge S and Remacle F 2015 {\em Physical Review B\/} {\bf 92}
  195302

\bibitem{Gamble-2015}
Gamble J~K, Jacobson N~T, Nielsen E, Baczewski A~D, Moussa J~E, Montano I and
  Muller R~P 2015 {\em Physical Review B\/} {\bf 91} 235318

\bibitem{Saraiva-2015}
Saraiva A~L, Baena A, Calder\'{o}n M~J and Koiller B 2015 {\em J. Phys.:
  Condens. Matter\/} {\bf 27} 154208

\bibitem{Urdampilleta-2015}
Urdampilleta M, Chatterjee A, Lo C~C, Kobayashi T, Mansir J, Barraud S, Betz
  A~C, Rogge S, Gonzalez-Zalba M~F and Morton J~J~L 2015 {\em Physical Review
  X\/} {\bf 5} 031024

\bibitem{Pica-2016}
Pica G, Lovett B~W, Bhatt R~N, Schenkel T and Lyon S~A 2016 {\em Physical
  Review B\/} {\bf 93} 035306

\bibitem{Harvey-2017}
Harvey-Collard P, Jacobson N~T, Rudolph M, Dominguez J, Eyck G~A~T, Wendt J~R,
  Pluym T, Gamble J~K, Lilly M~P, Pioro-Ladri\`ere M and Carroll M~S 2017 {\em
  Nature Communications\/} {\bf 8} 1029

\bibitem{Kane-1998}
Kane B~E 1998 {\em Nature\/} {\bf 393} 133

\bibitem{Loss-1998}
Loss D and DiVincenzo D~P 1998 {\em Physical Review A\/} {\bf 57} 120

\bibitem{DiVincenzo-2000}
DiVincenzo D~P, Bacon D, Kempe J, Burkard G,  and Whaley K~B 2000 {\em Nature
  (London)\/} {\bf 408} 339

\bibitem{Taylor-2005}
Taylor J, Engel H~A, D\"ur W, Yacoby A, Marcus C, Zoller P and Lukin M 2005
  {\em Nature Physics\/} {\bf 1} 177

\bibitem{Laird-2010}
Laird E, Taylor J, DiVincenzo D, Marcus C, Hanson M and Gossard A 2010 {\em
  Physical Review B\/} {\bf 82} 075403

\bibitem{Levy-2002}
Levy J 2002 {\em Physical Review Letters\/} {\bf 89} 147902

\bibitem{Petta-2005}
Petta J~R, C A, Taylor J~M, Laird E~A, Yacoby A, Lukin M~D, Marcus C~M, Hanson
  M~P and Gossard A~C 2005 {\em Science\/} {\bf 309} 2180

\bibitem{Pla-2013}
Pla J~J, Tan K~Y, Dehollain J~P, Lim W~H, Morton J~J~L, Zwanenburg F~A,
  Jamieson D~N, Dzurak A~S and Morello A 2013 {\em Nature\/} {\bf 334} 496

\bibitem{Ciorga-2001}
Ciorga M, Sachrajda A~S, Hawrylak P, Gould C, Zawadzki P, YFeng and ZWasilewski
  2001 {\em Physica E\/} {\bf 11} 35--40

\bibitem{Elzerman-2004}
Elzerman J~M, Hanson R, van Beveren L~H~W, Witkamp B, Vandersypen L~M~K and
  Kouwenhoven L~P 2004 {\em Nature\/} {\bf 430} 431

\bibitem{Gonzalez-2015}
Gonzalez-Zalba M, Barraud S, Ferguson A and Betz A 2015 {\em Nature
  Communications\/} {\bf 6} 6084

\bibitem{Li-2012}
Li R, Hu X and You J~Q 2012 {\em Physical Review B\/} {\bf 86} 205306

\bibitem{Zwanenburg-2013}
Zwanenburg F~A, Dzurak A~S, Morello A, Simmons M~Y, Hollenberg L~C~L, Klimeck
  G, Rogge S, Coppersmith S~N and Eriksson M~A 2013 {\em Rev. Mod. Phys.\/}
  {\bf 85} 961

\bibitem{XianWu-2014}
Wu X, Ward D~R, Prance J~R, Kim D, Gamble J~K, Mohr R~T, Shi Z, Savage D~E,
  Lagally M~G, Friesen M, Coppersmith S~N and Eriksson M~A 2014 {\em PNAS\/}
  {\bf 111} 11938 -- 11942

\bibitem{Barnes-2016}
Barnes E, Rudner M~S, Martins F, Malinowski F~K, Marcus C~M and Kuemmeth F 2016
  {\em Physical Review B\/} {\bf 93} 121407(R)

\bibitem{Pioro-2008}
Pioro-Ladri\`ere M, Obata T, Tokura Y, Shin Y~S, Kubo T, Yoshida K, Taniyama T
  and Tarucha S 2008 {\em Nature Physics\/} {\bf 4} 776--779

\bibitem{Shi-2012}
Shi Z, Simmons C~B, Prance J~R, Gamble J~K, Koh T~S, Shim Y~P, Hu X, Savage
  D~E, Lagally M~G, Eriksson M~A, Friesen M and Coppersmith S~N 2012 {\em
  Physical Review Letters\/} {\bf 108} 140503

\bibitem{Ferraro-2014}
Ferraro E, De~Michielis M, Mazzeo G, Fanciulli M and Prati E 2014 {\em Quantum
  Information Processing\/} {\bf 13} 1155--1173

\bibitem{Ferraro-2015-qip}
Ferraro E, De~Michielis M, Fanciulli M and Prati E 2015 {\em Quantum
  Information Processing\/} {\bf 14} 47--65

\bibitem{DeMichielis-2015}
De~Michielis M, Ferraro E, Fanciulli M and Prati E 2015 {\em Journal of Physics
  A: Mathematical and Theoretical\/} {\bf 48} 065304

\bibitem{Ferraro-2015-prb}
Ferraro E, De~Michielis M, Fanciulli M and Prati E 2015 {\em Physical Review
  B\/} {\bf 91} 075435

\bibitem{Ferraro-2017}
Ferraro E, Fanciulli M and {De Michielis} M 2017 {\em Quantum Information
  Processing\/} {\bf 16} 277

\bibitem{Ferraro-QIP2018}
Ferraro E, Fanciulli M and {De Michielis} M 2018 {\em Quantum Information
  Processing\/} {\bf 17} 130

\bibitem{Rotta-2016}
Rotta D, De~Michielis M, Ferraro E, Fanciulli M and Prati E 2016 {\em Quantum
  Information Processing Topical Collection\/} {\bf 15} 2253--2274

\bibitem{Koh-2012}
Koh T~S, Gamble J~K, Friesen M, Eriksson M~A and Coppersmith S~N 2012 {\em
  Physical Review Letters\/} {\bf 109} 250503

\bibitem{Kim-2012}
Kim D, Shi Z, Simmons C~B, Ward D~R, Prance J~R, Koh T~S, Gamble J~K, Savage
  D~E, Lagally M~G, Friesen M, Coppersmith S~N and Eriksson M~A 2014 {\em
  Nature\/} {\bf 511} 70--74

\bibitem{Kim-2015}
Kim D, Ward D~R, Simmons C~B, Savage D~E, Lagally M~G, Friesen M, Coppersmith
  S~N and Eriksson M~A 2015 {\em Npj Quantum Information\/} {\bf 1} 15004

\bibitem{Thorgrimsson-2017}
Thorgrimsson B, Kim D, Yang Y~C, Smith L~W, Simmons C~B, Ward D~R, Foote R~H,
  Corrigan J, Savage D~E, Lagally M~G, Friesen M, Coppersmith S~N and Eriksson
  M~A 2017 {\em Npj Quantum Information\/} {\bf 3} 32

\bibitem{Mohammady-2012}
Mohammady M~H, Morley G~W, Nazir A and Monteiro T~S 2012 {\em Physical Review
  B\/} {\bf 85} 094404

\bibitem{Sousa-2003}
de~Sousa R and Sarma S~D 2003 {\em Physical Review B\/} {\bf 68} 115322

\bibitem{Wolfowicz-2013}
Wolfowicz G, AlexeiMTyryshkin, George R~E, Riemann H, Abrosimov N~V, Becker P,
  Pohl H~J, Thewalt M~L~W, Lyon S~A and Morton J~J~L 2013 {\em Nature
  Nanotechnology\/} {\bf 8} 561

\bibitem{Morello-2010}
Morello A, Pla J~J, Zwanenburg F~A, Chan K~W, Tan K~Y, Huebl H, M\"ott\"onen M,
  Nugroho C~D, Yang C, van Donkelaar J~A, Alves A~D~C, Jamieson D~N, Escott
  C~C, Hollenberg L~C~L, Clark R~G and Dzurak A~S 2010 {\em Nature\/} {\bf 467}
  687

\bibitem{Muhonen-2014}
Muhonen J~T, Dehollain J~P, Laucht A, Hudson F~E, Kalra R, Sekiguchi T, Itoh
  K~M, Jamieson D~N, McCallum J~C, Dzurak A~S and Morello A 2014 {\em Nature
  Nanotechnology\/} {\bf 9} 986

\bibitem{Zajac-2018}
Zajac D~M, Sigillito A~J, Russ M, Borjans F, Taylor J~M, Burkard G and Petta
  J~R 2018 {\em Science\/} {\bf 359} 439--442

\bibitem{Li-2018}
Li R, Petit L, Franke D~P, Dehollain J~P, Helsen J, Steudtner M, Thomas N~K,
  Yoscovits Z~R, Singh K~J, Wehner S, Vandersypen L~M~K, Clarke J~S and
  Veldhorst M 2018 {\em Science Advances\/} {\bf 4} eaar3960

\bibitem{Laucht-2015}
Laucht A, Muhonen J~T, Mohiyaddin F~A, Kalra R, Dehollain J~P, Freer S, Hudson
  F~E, Veldhorst M, Rahman R, Klimeck G, Itoh K~M, Jamieson D~N, McCallum J~C,
  Dzurak A~S and Morello A 2015 {\em Science Advances\/} {\bf 1} e1500022

\bibitem{Nielsen-2000}
Nielsen M~A and Chuang I~L 2000 {\em Quantum Computation and Quantum
  Information\/} (Cambridge: Cambridge University Press)

\bibitem{Marinescu-2012}
Marinescu D~C and Marinescu G~M 2012 {\em Classical and Quantum Information\/}
  (Amsterdam: Elsevier)

\bibitem{DatasheetMWgenerator}
Keysight T 2017 {\em Keysight Technologies E8257D PSG Microwave Analog Signal
  Generator\/}

\bibitem{Khaneja-2005}
Khaneja N, Reiss T, Kehlet C, Schulte-Herbr\"uggen T and Glaser S~J 2005 {\em
  J. Magn. Reson.\/} {\bf 172} 296

\end{thebibliography}

\end{document}